\documentclass[12pt]{article}
\usepackage{graphicx,amsmath,amssymb}
\usepackage{epsfig}
\allowdisplaybreaks

\title{\bf Long Distance Effects in Mixed Electromagnetic-Gravitational Scattering}
\author{Barry R. Holstein$^a$\footnote{\tt holstein@physics.umass.edu} \hspace*{1pt} and Andreas
Ross$^{a,b}$\footnote{\tt
andreas.ross@yale.edu} \\ \\
$^a$ Department of Physics -- LGRT\\
University of Massachusetts\\
Amherst, MA  01003, USA\\  \\
$^b$ Department of Physics \\
 Yale University \\
New Haven, CT 06520, USA \\ \\}

\begin{document}
\maketitle
\thispagestyle{empty}

\begin{abstract}
Using the methods of effective field theory we examine
long range effects in mixed electromagnetic-gravitational
scattering. Recent calculations which have yielded
differing results for such effects are examined and corrected.
We consider various spin configurations of the scattered particles
and find that universality with respect to spin-dependence is obtained in
agreement with expectations.
\end{abstract}

\vspace{0.2 in}
\setcounter{page}{0}
\newpage

\section{Introduction}

Recently there have been a number of calculations of long distance
effects in mixed electromagnetic-gravitational scattering ({\it
i.e.}, $\mathcal O(G \alpha)$ effects), using the methods of
effective field theory \cite{bjb, msb, svf}. The basic idea here is
that such long distance corrections arise from pieces of the
scattering amplitude which are nonanalytic in the momentum transfer.
Such nonanalytic components can be isolated from one loop scattering
amplitudes that involve both photons and gravitons using effective
field theory methods, a procedure pioneered more than a decade ago
by Donoghue in the case of the gravitational interaction \cite{don}.
There are two distinct forms of such corrections---a classical piece
(independent of $\hbar$) which arises from square root components
$\sim 1/\sqrt{-q^2}$ and a quantum ($\hbar$-dependent) component
associated with logarithmic nonanalyticity $\sim \log -q^2$ where q
is the four-momentum transfer in the scattering reaction.

The first such calculation was done by N.E.J. Bjerrum-Bohr in the case of the scattering of
two nonidentical spinless particles, particle $a$ of mass $m_a$ and charge $Z_a |e|$ and
particle $b$ of mass $m_b$ and charge $Z_b |e|$, who quoted the one loop
potential \cite{bjb}\footnote{Note, however, this result has now been corrected by Bjerrum-Bohr \cite{nb2}.}
\begin{eqnarray}
{}^0V_{CG}^{(2)}(\vec{r})&=&G\alpha\left[{1\over
2}{m_aZ_b^2+m_bZ_a^2\over
r^2}+3{Z_aZ_b(m_a+m_b)\over r^2}\right.\nonumber\\
&& {} \hspace*{18pt} -\left.{4\hbar\over 3\pi r^3}\left(Z_b^2{m_a\over
m_b}+Z_a^2{m_b\over m_a}\right)-{8Z_aZ_b\hbar\over \pi r^3}\right]
\end{eqnarray}
A followup calculation by Butt involving the spin-averaged
spin-1/2 -- spin-1/2 scattering amplitude determined the
form \cite{msb}
\begin{eqnarray}
\left<{}^{{1\over 2}{1\over 2}}V_{CG}^{(2)}(\vec{r})\right>_{spin-av.}&=&G\alpha\left[{1\over
2}{m_aZ_b^2+m_bZ_a^2\over
r^2}+3{Z_aZ_b(m_a+m_b)\over r^2}\right.\nonumber\\
&& {} \hspace*{18pt} -\left.{4\hbar\over 3\pi
r^3}\left(Z_b^2{m_a\over m_b}+Z_a^2{m_b\over m_a}\right)-
{\mbox{$\frac{15}{6}$}} {Z_aZ_b\hbar\over \pi r^3}\right].
\end{eqnarray}
What is surprising here is that while the classical ($\sim 1/r^2)$
pieces of the spin-0 -- spin-0 and of the spin-averaged spin-1/2 --
spin-1/2 potentials agree, the quantum-mechanical ($\sim \hbar/r^3)$
components have differing forms. These results then constitute a
deviation from the universality of similar second order potentials
found in the case of purely electromagnetic and purely gravitational
scattering in recent calculations by Holstein and Ross \cite{hrem,
hrgr}. In the most recent calculation of this $\mathcal O(G \alpha)$
potential in the case of spinless scattering by Faller, it is
claimed to have still a third form \cite{svf}
\begin{eqnarray}
{}^0V_{CG}^{(2)}(\vec{r})&=&G\alpha\left[{1\over
2}{m_aZ_b^2+m_bZ_a^2\over
r^2}+3{Z_aZ_b(m_a+m_b)\over r^2}\right.\nonumber\\
&& {} \hspace*{18pt} - \left.{4\hbar\over 3\pi r^3}\left(Z_b^2{m_a\over
m_b}+Z_a^2{m_b\over m_a}\right)-{2Z_aZ_b\hbar\over 3\pi r^3}\right].
\end{eqnarray}

The surprising disagreement between these results and the possible
breakdown of universality indicated by the discrepancy between them
clearly calls for a new evaluation which resolves previous
disagreements and clearly answers the question of universality. This
is the purpose of the work below.  We shall evaluate mixed gravitational-electromagnetic
scattering at the one loop level in the case of
spin-0 -- spin-0, spin-0 -- spin-1/2, spin-0 -- spin-1,
and spin-1/2 -- spin-1/2 scattering, looking both at the
spin-independent and spin-dependent pieces of the scattering
amplitude.  We quote the contributions of each diagram in order to
assess and resolve the problems with previous evaluations.

In the next section we perform the calculation for the case of
spinless scattering and compare with the previous evaluations by
Bjerrum-Bohr \cite{bjb} and Faller \cite{svf}.  Then in the following
sections we generalize this calculation to the case wherein one or
both particles carry spin, connecting with the calculation by
Butt \cite{msb}. Our results, which clearly resolve the questions
raised above, are summarized in a concluding section. While
some of our calculational methods are found in the appendix, we refer
to our companion papers on purely electromagnetic scattering \cite{hrem}
and on purely gravitational scattering \cite{hrgr} for much of the
calculational ingredients such as the loop integrals and parts of the
Feynman rules needed.

\section{Spin-0 -- Spin-0 Scattering}

We begin our discussion by considering the case of the scattering of
two spinless particles, particle $a$ of mass $m_a$ and charge $Z_a |e|$ and
particle $b$ of mass $m_b$ and charge $Z_b |e|$.  We choose the initial (final)
four-momentum of particle $a$ to be $p_1$ ($p_2=p_1-q$) and that of
particle $b$ to be $p_3$ ($p_4=p_3+q$).  This is the case originally
studied by Bjerrum-Bohr \cite{bjb}.

\begin{figure}[t]
\begin{center}
\epsfig{file=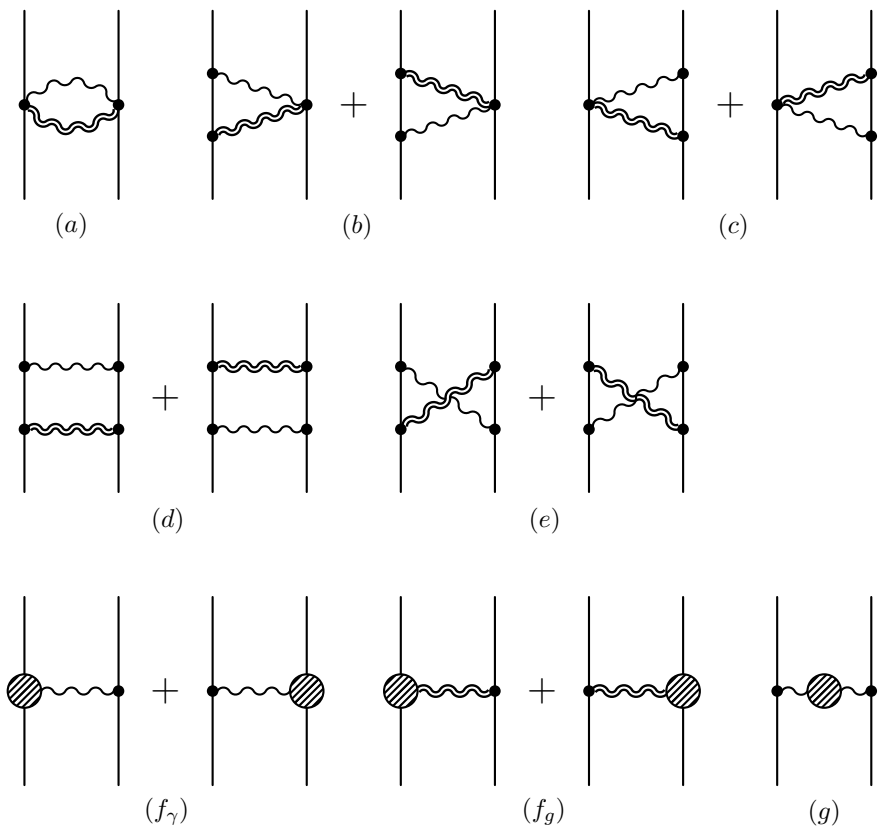,width=0.85\textwidth} \caption{One loop
diagrams in mixed electromagnetic-gravitational scattering. } \label{fig_diags}
\end{center}
\end{figure}

\begin{figure}[t]
 \begin{center}
  \epsfig{file=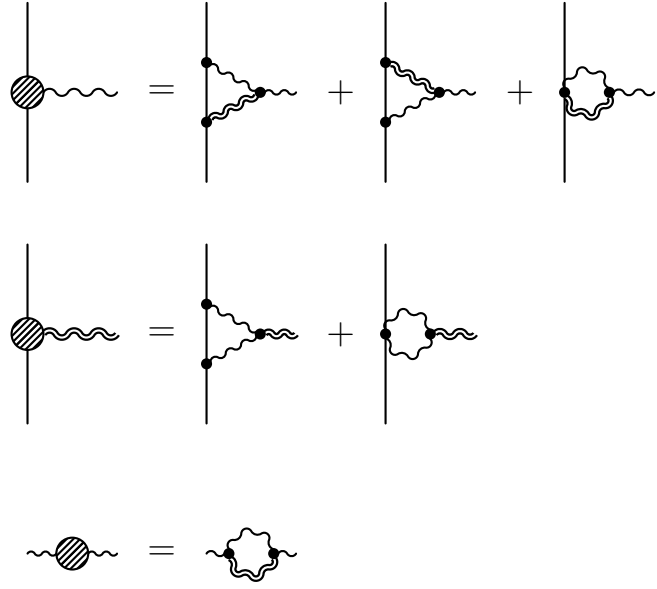,width=0.63\textwidth}
  \caption{Loop corrections subsumed in vertex and in vacuum polarization functions for mixed electromagnetic-gravitational scattering.}
  \label{fig_blobs}
\end{center}
\end{figure}

The corresponding one loop diagrams that give rise to long distance
contributions are drawn in Fig. \ref{fig_diags} where the blobs are
explained in Fig. \ref{fig_blobs}. Since we will be comparing
different calculations, it is useful to present the amplitudes for
each diagram separately. The calculational techniques including the
Feynman rules and the loop integrals needed are outlined in
\cite{hrem}, \cite{hrgr} and in Appendix \ref{app_fmr}. Defining the
nonanalytic structures
$$L=\log-q^2\qquad{\rm and}\qquad S={\pi^2\over \sqrt{-q^2}}$$
we find for the case of spinless scattering the contributions

\begin{eqnarray}
{}^0{\cal M}_{\ref{fig_diags}a}^{(2)}(q) \hspace*{-3pt} &=&\hspace*{-3pt} G \alpha Z_a Z_b \left[16 L\right]\nonumber\\
{}^0{\cal M}_{\ref{fig_diags}b}^{(2)}(q) \hspace*{-3pt} &=&\hspace*{-3pt} G \alpha Z_a Z_b \left[-16 L - 8 m_a S\right]\nonumber\\
{}^0{\cal M}_{\ref{fig_diags}c}^{(2)}(q) \hspace*{-3pt} &=&\hspace*{-3pt} G \alpha Z_a Z_b \left[-16 L - 8 m_b S\right]\nonumber\\
{}^0{\cal M}_{\ref{fig_diags}d}^{(2)}(q) \hspace*{-3pt} &=&\hspace*{-3pt} G \alpha Z_a Z_b \left[\hspace*{-0.5pt} \left( \hspace*{-1pt} - \frac{8 m_a m_b}{q^2} - \frac{7 (m_a^2 + m_b^2)}{m_a m_b} + 4 \right) \hspace*{-1.5pt} L - \hspace*{-0.5pt}5 (m_a + m_b) S \right]\nonumber\\
&+& \hspace*{-5pt} i 8\pi G  m_a m_b \, \alpha Z_a Z_b \hspace*{1.5pt} {L\over q^2} \sqrt{m_a m_b \over s-s_0}\nonumber\\
{}^0{\cal M}_{\ref{fig_diags}e}^{(2)}(q) \hspace*{-3pt} &=&\hspace*{-3pt} G \alpha Z_a Z_b \left[\hspace*{-0.5pt} \left( \hspace*{-1pt} \frac{8 m_a m_b}{q^2} + \frac{7 (m_a^2 + m_b^2)}{m_a m_b} + \frac{68}{3} \right) \hspace*{-1.5pt} L + \hspace*{-0.5pt}5 (m_a + m_b) S \right]\nonumber\\
{}^0{\cal M}_{\ref{fig_diags}f_\gamma}(q) \hspace*{-3pt} &=&\hspace*{-3pt} G \alpha Z_a Z_b \left[2 (m_a + m_b) S\right]\nonumber\\
{}^0{\cal M}_{\ref{fig_diags}f_g}(q) \hspace*{-3pt} &=&\hspace*{-3pt} G \alpha \left[- \frac{8(Z_a^2 m_b^2 + Z_b^2 m_a^2)}{3 m_a m_b} \hspace*{1pt} L  - (Z_a^2 m_b + Z_b^2 m_a) S\right]\nonumber\\
{}^0{\cal M}_{\ref{fig_diags}g}(q) \hspace*{-3pt} &=&\hspace*{-3pt} G \alpha Z_a Z_b \left[{4 \over 3} L \right].
\end{eqnarray}
from each diagram---(a)-(g) in Fig. \ref{fig_diags}---where $s=(p_1+p_3)^2$ is the
square of the center of mass energy and $s_0=(m_a+m_b)^2$ is its
threshold value\footnote{Our
normalization of the amplitudes is a nonrelativistic one such that
after applying all Feynman rules we divide the amplitude by a factor
of $\sqrt{2E_12E_22E_32E_4}$.}.

Comparing with previous work, we agree diagram by diagram with the
calculation of Bjerrum-Bohr.  (In the case of the paper by Faller,
we are also in agreement---once certain typos are corrected---except
for the box and cross-box diagrams, which are too small in his
calculation by a factor of two.)  When all contributions are added
together, we find the total amplitude
\begin{eqnarray}
{}^0{\cal M}^{(2)}_{tot}(q) \hspace*{-5pt} & = \hspace*{-5pt} &G \alpha \bigg[Z_a Z_b (-6 (m_a+m_b)S + 12 L) \nonumber \\
&& \hspace*{13.5pt} {} - (Z_a^2 m_b + Z_b^2 m_a) S - \frac{8(Z_a^2 m_b^2 + Z_b^2 m_a^2)}{3 m_a m_b} \, L\bigg] \nonumber \\
&+& \hspace*{-5pt} i 8\pi G  m_a m_b \, \alpha Z_a Z_b \hspace*{1.5pt} {L\over q^2} \sqrt{m_a m_b \over s-s_0}. \label{eq_amp_00}
\end{eqnarray}
For later use we shall work in the nonrelativistic limit and in the
center of mass frame---$\vec{p}_1+\vec{p}_3=0$.  We have then
\begin{equation}
s-s_0=2\sqrt{m_a^2+\vec p_1^{\hspace*{1.4pt} 2}}\sqrt{m_b^2+\vec
p_1^{\hspace*{1.4pt} 2}}+2\vec p_1^{\hspace*{1.4pt} 2}-2m_am_b \label{eq_sms0_nonrel}
\end{equation}
and
\begin{equation}
\sqrt{m_am_b\over s-s_0}\simeq {m_r\over p_0} \label{eq_rootsms0_nonrel}
\end{equation}
where $m_r=m_am_b/(m_a+m_b)$ is the reduced mass and $p_0 \equiv
|\vec{p}_i|,\quad i=1,2,3,4$. The total amplitude then becomes
\begin{eqnarray}
{}^0{\cal M}^{(2)}_{tot}(\vec q) \hspace*{-5pt} & \simeq \hspace*{-5pt} & G \alpha \bigg[Z_a Z_b (-6 (m_a+m_b)S + 12 L) \nonumber \\
&& \hspace*{13.5pt} {} - (Z_a^2 m_b + Z_b^2 m_a) S - \frac{8(Z_a^2 m_b^2 + Z_b^2 m_a^2)}{3 m_a m_b} \, L\bigg] \nonumber \\
&+& \hspace*{-5pt} i 8\pi G  m_a m_b \, \alpha Z_a Z_b \hspace*{1.5pt} {L\over q^2} \frac{m_r}{p_0}. \label{eq_amp_00_nr}
\end{eqnarray}

At this point we encounter a difficulty in defining a proper second
order potential in that the inclusion of loop effects has produced
an imaginary piece of the transition amplitude, which clearly cannot
be part of a properly defined (hermitian) potential.  Of course,
there is no mystery as to why this term is present, since unitarity
{\it requires} the presence of such imaginary components. The
solution is also clear. We must subtract from the second order
transition amplitude the piece which arises from the iterated lowest
order potential before attempting to identify a proper second order
potential. Indeed, in lowest order spinless scattering, we use the
Newtonian potential to describe the gravitational potential
\begin{equation}
{}^0V_G^{(1)}(\vec r)=-G{m_am_b\over r}
\end{equation}
while for its electromagnetic analog we utilize the Coulomb
potential
\begin{equation}
{}^0V_C^{(1)}(\vec r)={Z_aZ_b\alpha\over r}.
\end{equation}
Working in momentum space we have
\begin{eqnarray}
{}^0V_G^{(1)}(\vec {q}) & = & \left<\vec{p}_f \left|{}^0 \hat V_G^{(1)} \right|\vec{p}_i\right> = - \frac{4\pi Gm_am_b}{\vec q^{\hspace*{1.4pt} 2}} = -{4\pi Gm_am_b\over (\vec{p}_i-\vec{p}_f)^2} \\
{}^0V_C^{(1)}(\vec {q}) & = & \left<\vec{p}_f \left|{}^0 \hat V_C^{(1)} \right|\vec{p}_i\right> = \frac{4\pi \alpha Z_aZ_b}{\vec q^{\hspace*{1.4pt} 2}} = {4\pi \alpha Z_aZ_b\over (\vec{p}_i-\vec{p}_f)^2}
\end{eqnarray}
respectively.  The relevant second Born term of $\mathcal O(G
\alpha)$ is then\footnote{Note that we elect to use here the simple
nonrelativistic forms for both the potentials and the propagator.
We shall comment later on this choice.}
\begin{eqnarray} \label{eq:iteration00a}
{}^0{\rm Amp}_{CG}^{(2)}(\vec q)&=&- \int{d^3\ell\over
(2\pi)^3} \, \frac{\left<\vec p_f \left| {}^0 \hat V^{(1)}_G \right| \vec \ell \, \right> \left<\vec \ell \left| {}^0 \hat V^{(1)}_C \right| \vec p_i \right>}{E(p_0) - E(\ell) + i \epsilon}\nonumber\\
&& - \int{d^3\ell\over
(2\pi)^3} \, \frac{\left<\vec p_f \left| {}^0 \hat V^{(1)}_C \right| \vec \ell \, \right> \left<\vec \ell \left| {}^0 \hat V^{(1)}_G \right| \vec p_i \right>}{E(p_0) - E(\ell) + i \epsilon}\nonumber\\
&=&i\int{d^3\ell\over (2\pi)^3}{}^0V_G^{(1)}(\vec \ell - \vec p_f)
\, G^{(0)}(\vec{\ell}) \, {}^0V_C^{(1)}(\vec{p}_i-\vec{\ell} \, ) \\
&+&i\int{d^3\ell\over (2\pi)^3}{}^0V_C^{(1)}(\vec \ell - \vec p_f)
\, G^{(0)}(\vec{\ell}) \, {}^0V_G^{(1)}(\vec{p}_i-\vec{\ell} \, )
\end{eqnarray}
where
\begin{equation}
G^{(0)}(\ell)={i\over {p_0^2\over 2m_r}-{\ell^2\over
2m_r}+i\epsilon}
\end{equation}
is the free propagator.  The remaining integration can be performed
exactly, as discussed in Appendix A, by including a photon "mass"
term $\lambda^2$ as a regulator. Identifying $\vec p_i$ with $\vec p_1$ and $\vec p_f$ with $\vec   p_2$, the iteration amplitude reads
\begin{eqnarray}
{}^0{\rm Amp}_{CG}^{(2)}(\vec q) \hspace*{-5pt} &= \hspace*{-5pt} &i\int {d^3\ell\over (2\pi)^3} \left[{-4\pi
Gm_am_b\over |\vec{\ell} - \vec{p}_2|^2+\lambda^2}{i\over {p_0^2\over
2m_r}-{\ell^2\over 2m_r}+i\epsilon}{Z_aZ_be^2\over
|\vec{p}_1 - \vec{\ell} \hspace*{1pt}|^2+\lambda^2} \right.\nonumber\\
&& \hspace*{48pt} \left. + {Z_aZ_be^2\over |\vec{\ell} - \vec{p}_2|^2+\lambda^2}{i\over
{p_0^2\over 2m_r}-{\ell^2\over 2m_r}+i\epsilon}{-4\pi Gm_am_b\over
|\vec{p}_1 - \vec{\ell} \hspace*{1pt}|^2+\lambda^2} \right] \nonumber\\
& \stackrel{\lambda\rightarrow 0}{\longrightarrow}& \hspace*{-4pt} 2H
  = i 8\pi G  m_a m_b \, \alpha Z_a Z_b \hspace*{1.5pt} {L\over q^2} \frac{m_r}{p_0}
\end{eqnarray}
which precisely reproduces the imaginary component of ${}^0{\cal
M}_{tot}^{(2)}(\vec{q})$, as expected.  In order to generate a
properly defined second order potential ${}^0V^{(2)}_{CG}(\vec{r})$
we must subtract this second order Born term from the second order
amplitude, yielding the result
\begin{eqnarray}
{}^0V_{CG}^{(2)}(\vec{r})&=&-\int{d^3q\over
(2\pi)^3}e^{-i\vec{q}\cdot\vec{r}}\left[{}^0{\cal
M}_{tot}^{(2)}(\vec q)-{}^0{\rm Amp}_{CG}^{(2)}(\vec q)\right]\nonumber\\
&=& G\alpha\left[{1\over 2}{m_aZ_b^2+m_bZ_a^2\over
r^2}+3{Z_aZ_b(m_a+m_b)\over r^2}\right.\nonumber\\
&& {} \hspace*{18pt} -\left.{4\hbar\over 3\pi r^3}\left(Z_b^2{m_a\over
m_b}+Z_a^2{m_b\over m_a}\right)+{6Z_aZ_b\hbar\over \pi
r^3}\right]\quad\label{eq:so}
\end{eqnarray}
which disagrees with that quoted by both Bjerrum-Bohr and Faller
\cite{bjb, svf}. In the case of the Bjerrum-Bohr this is due to a
typo which occurred in his total result. This mistake has been
corrected in an erratum \cite{nb2}, so that our results are now in
agreement. In the case of Faller the disagreement is due to his use
of box plus cross-box contributions which are a factor of two
smaller than the correct form. We notice that the resulting
$\mathcal O(G \alpha)$ potential in Eq. (\ref{eq:so}) displays two
different dependences on the charges of the scattered particles: 
The terms proportional to $Z_a Z_b$ which vanish once one of the scattered particles
has zero charge and the terms proportional to $Z_a^2$ and $Z_b^2$ which can persist 
if one of the scattered particles is uncharged. Thus we avoid the temptation to call 
our results gravitational corrections to electromagnetic scattering
since the $Z_a^2$ and $Z_b^2$ terms cannot be viewed as such
corrections.

It should be noted here that the form of the classical
($\hbar$-independent) component of the second order potential is
ambiguous, as pointed out by Sucher \cite{jsu}. For a detailed
discussion of these ambiguities in the case of purely
electromagnetic scattering, see \cite{hrem}. The point is that the
form of the iterated lowest order potentials depends upon the
precise form of the lowest potentials and whether relativistic
effects are included in it and in the propagator $G^{(0)}(\ell)$.
The existence of such ambiguity is not a problem, since the
potential is {\it not} an observable.  What {\it is} an observable
and {\it is} invariant is the second order on-shell transition amplitude,
which is given by
\begin{equation}
{}^0{\cal M}_{tot}^{(2)}(\vec{q})={}^0{\rm Amp}^{(2)}(\vec{q})-\int
d^3r \, e^{i\vec{q}\cdot\vec{r}} \ {}^0V_{CG}^{(2)}(\vec{r}).
\end{equation}
Because of this invariance, we shall for simplicity utilize the
simple nonrelativistic potentials and propagator in obtaining the
iterated second order amplitude.

\section{Spin-0 -- Spin-1/2 Scattering}

Before proceeding to the spin-1/2 -- spin-1/2 calculation of Butt,
we first examine the simpler case of a spin-1/2 particle $b$
scattering from a spinless particle $a$, in order to check the
hypothesis of universality.  The calculation proceeds as in the
previous section except that for the vertices for particle $b$ we
use the spin-1/2 forms. One other subtlety that arises in the
calculation once spin is involved is that it contains {\it two}
independent kinematic variables: the momentum transfer $q^2$ and $s
- s_0$, which is to leading order proportional to $p_0^2$ (where
$p_0^2 \equiv \vec p_i^{\hspace*{1.4pt}2},\quad i=1,2,3,4$) in the
center of mass frame.  We find that our results differ if we perform
an expansion first in $s - s_0$ and then in $q^2$ or vice versa.
This ordering issue only occurs for the box diagram, diagram (d) of
Fig. \ref{fig_diags}, where it stems from the reduction of vector
and tensor box integrals. Their reduction in terms of scalar
integrals involves the inversion of a matrix whose Gram determinant
vanishes in the nonrelativistic threshold limit $q^2, s - s_0
\rightarrow 0$. More precisely, the denominators or the vector and
tensor box integrals involve a factor of $(4 p_0^2 - \vec
q^{\hspace*{1.4pt} 2})$ when expanded in the nonrelativistic limit.
Since $q^{\hspace*{1.4pt} 2} = 4 p_0^2 \sin^2 \frac{\theta}{2}$ with
$\theta$ the scattering angle, we notice that $4 p_0^2 > \vec
q^{\hspace*{1.4pt} 2}$ unless we consider backward scattering where
$\theta = \pi$ and where the scattering amplitude diverges. And
since $p_0^2$ originates from the relativistic structure $s - s_0$,
we therefore must first expand our vector and tensor box integrals
in $q^2$ and then in $s - s_0$. In this way we find
\begin{eqnarray}
{}^{1\over 2}{\cal M}_{\ref{fig_diags}a}^{(2)}(q) \hspace*{-5pt} &=\hspace*{-5pt}&G \alpha Z_a Z_b \bigg[{1\over
m_a}\bar{u}(p_4) \! \not \!{p}_1u(p_3) \, 6L \bigg]\nonumber\\
{}^{1\over 2}{\cal M}_{\ref{fig_diags}b}^{(2)}(q) \hspace*{-5pt} &=\hspace*{-5pt}&G \alpha Z_a Z_b \bigg[{1\over m_a}\bar{u}(p_4) \! \not \!{p}_1u(p_3) (-6 m_a S -9 L ) \bigg]\nonumber\\
{}^{1\over 2}{\cal M}_{\ref{fig_diags}c}^{(2)}(q)  \hspace*{-5pt} &=\hspace*{-5pt}&G \alpha Z_a Z_b \bigg[\bar{u}(p_4) u(p_3) (- 6 m_b S - 4 L) \nonumber\\
&& \hspace*{41pt} + {1\over m_a}\bar{u}(p_4) \! \not \!{p}_1u(p_3) (- 2 m_b S - 2 L ) \bigg]\nonumber\\
{}^{1\over 2}{\cal M}_{\ref{fig_diags}d}^{(2)}(q)  \hspace*{-5pt} &=\hspace*{-5pt}&G \alpha Z_a Z_b \Bigg[\bar{u}(p_4)u(p_3) \hspace*{-1pt} \Bigg( \frac{4 m_a m_b}{q^2} L + {2 m_am_b(2 m_a+3 m_b)\over s-s_0}S \nonumber\\
&&  \hspace*{98pt} + \hspace*{0.9pt} (3 m_a \hspace*{-2pt} + \hspace*{-2pt} 8 m_b) S  \hspace*{-1pt} +  \hspace*{-1pt} \frac{10 m_a^2 \hspace*{-2pt} + \hspace*{-2pt} 10m_am_b \hspace*{-2pt} - \hspace*{-2pt} 3 m_b^2}{3m_am_b} L \hspace*{-1pt}\Bigg)\nonumber\\
&& \hspace*{40pt} + {1\over m_a}\bar{u}(p_4\hspace*{-0.4pt}) \hspace*{-1.2pt} \! \not \!{p}_1u(p_3\hspace*{-0.4pt}) \hspace*{-1pt}\Bigg(\! \! \hspace*{-1.2pt} - \! \frac{12m_am_b}{q^2} L \hspace*{-1pt}
- \hspace*{-2pt} {2 m_am_b(2 m_a \hspace*{-2.9pt} + \hspace*{-2pt} 3 m_b)\over s-s_0} S\nonumber\\
&& \hspace*{133pt} -  (7 m_a + 10 m_b) S \nonumber\\
&& \hspace*{133pt} -  \frac{44 m_a^2 - m_a m_b + 36 m_b^2}{6 m_a m_b} L \Bigg) \Bigg]\nonumber\\
& + \hspace*{-5pt} & {} \hspace*{-0.9pt} i 8\pi G  m_a m_b \hspace*{0.5pt} \alpha Z_a Z_b \hspace*{0.8pt} {L\over q^2} \sqrt{m_a m_b \over s-s_0} \hspace*{-2.5pt} \left(\hspace*{-1.4pt} \frac{3}{2} \hspace*{0.6pt}  {1\over m_a}\bar{u}(p_4 \hspace*{-0.4pt}) \hspace*{-1.4pt} \! \not \!{p}_1 \hspace*{-0.3pt} u(p_3\hspace*{-0.4pt}) \hspace*{-2.3pt} - \hspace*{-2pt} \frac{1}{2}\bar{u}(p_4\hspace*{-0.4pt})u(p_3\hspace*{-0.4pt})\hspace*{-4.2pt}\right)\nonumber\\
{}^{1\over 2}{\cal M}_{\ref{fig_diags}e}^{(2)}(q)  \hspace*{-5pt} &=\hspace*{-5pt}&G \alpha Z_a Z_b \Bigg[\bar{u}(p_4)u(p_3) \hspace*{-1pt} \Bigg( - \frac{4 m_a m_b}{q^2} L\nonumber\\
&&  \hspace*{99pt} - \frac{4 m_a \hspace*{-1pt} - \hspace*{-0.7pt} 3 m_b}{2} S - \hspace*{-1pt} \frac{10 m_a^2 \hspace*{-0.8pt} - \hspace*{-0.5pt} 6m_am_b \hspace*{-1pt} - \hspace*{-0.5pt} 3 m_b^2}{3m_am_b} L \hspace*{-1pt}\Bigg)\nonumber\\
&& \hspace*{41pt} + {1\over m_a}\bar{u}(p_4) \! \not \!{p}_1u(p_3)
\hspace*{-1pt}\Bigg({12m_am_b\over q^2} L + \frac{12 m_a + 7 m_b}{2} S \nonumber\\
&& \hspace*{134pt} +  \frac{44 m_a^2 + 85 m_a m_b + 36 m_b^2}{6 m_a m_b} L \Bigg) \Bigg]\nonumber\\
{}^{1\over 2}{\cal M}_{\ref{fig_diags}f_\gamma}^{(2)}(q) \hspace*{-5pt} &=\hspace*{-5pt}&G \alpha Z_a Z_b \bigg[\bar{u}(p_4) u(p_3) \, m_b S + {1\over m_a}\bar{u}(p_4) \! \not \!{p}_1u(p_3) \hspace*{1pt} (2 m_a + m_b) S \hspace*{0.8pt} \bigg]\nonumber\\
{}^{1\over 2}{\cal M}_{\ref{fig_diags}f_g}^{(2)}(q) \hspace*{-5pt} &=\hspace*{-5pt}&G \alpha \Bigg[\bar{u}(p_4) u(p_3) \bigg(\frac{Z_a^2 m_b}{2} \hspace*{1pt} S + \frac{4(Z_a^2 m_b^2 - Z_b^2 m_a^2)}{3 m_a m_b} \bigg)\nonumber\\
&& \hspace*{16pt} + {1\over m_a}\bar{u}(p_4) \! \not \!{p}_1u(p_3) \bigg(- \frac{3 Z_a^2 m_b + 2 Z_b^2 m_a}{2}S \nonumber \\
&& \hspace*{113pt} - \frac{4(3 Z_a^2 m_b^2 + Z_b^2 m_a^2)}{3 m_a m_b} L \bigg) \Bigg]\nonumber \\
{}^{1\over 2}{\cal M}_{\ref{fig_diags}g}^{(2)}(q) \hspace*{-5pt} &=\hspace*{-5pt}&G \alpha Z_a Z_b \bigg[{1\over m_a}\bar{u}(p_4) \! \not \!{p}_1u(p_3) \bigg( \frac{4}{3}L\bigg) \bigg].
\end{eqnarray}
where our spinors are normalized as $\bar u(p) \hspace*{1pt} u(p) =
1$.  Summing, we find the total transition amplitude
\begin{eqnarray}
{}^{1\over 2}{\cal M}_{tot}^{(2)}(q) \hspace*{-6pt} &=\hspace*{-6pt}&G \alpha Z_a Z_b \Bigg[ \hspace*{-0.8pt} S \hspace*{-0.4pt} \bigg(\hspace*{-3.8pt} (\hspace*{-0.4pt}m_a \hspace*{-2.6pt} + \hspace*{-2.1pt} 5 m_b\hspace*{-0.4pt}) \bar{u}(p_4\hspace*{-0.4pt})u(p_3\hspace*{-0.4pt}) \hspace*{-1.9pt} - \hspace*{-1.7pt}(\hspace*{-0.4pt}5 m_a \hspace*{-2.6pt} + \hspace*{-2pt}9 m_b\hspace*{-0.4pt}) \hspace*{0.2pt}  {1\over m_a} \bar{u}(p_4\hspace*{-0.4pt}) \hspace*{-1.5pt} \! \not\!{p}_1u(p_3\hspace*{-0.4pt})\hspace*{-3.5pt}\bigg)\nonumber\\
&&\hspace*{38pt}+ L\left(\frac{4}{3}\bar{u}(p_4)u(p_3) +{32\over 3}{1\over m_a}\bar{u}(p_4) \! \not\!{p}_1u(p_3)\right) \nonumber\\
&&\hspace*{38pt}+ {2 m_a m_b (2 m_a\hspace*{-1.9pt}+\hspace*{-1.4pt}3m_b) S\over
s-s_0}\hspace*{-1.5pt}\left(\hspace*{-1pt} \bar{u}(p_4)u(p_3) \hspace*{-1.3pt}-\hspace*{-1.3pt}{1\over
m_a}\bar{u}(p_4) \hspace*{-1.2pt} \! \not\!{p}_1u(p_3)\hspace*{-2pt}\right)\hspace*{-2.8pt}\Bigg]\nonumber\\
&+\hspace*{-6pt}&G \alpha \Bigg[\bar{u}(p_4) u(p_3) \bigg(\frac{Z_a^2 m_b}{2} \hspace*{1pt} S + \frac{4(Z_a^2 m_b^2 - Z_b^2 m_a^2)}{3 m_a m_b} \bigg)\nonumber\\
&& \hspace*{17pt} + {1\over m_a}\bar{u}(p_4) \! \not \!{p}_1u(p_3) \bigg(- \frac{3 Z_a^2 m_b + 2 Z_b^2 m_a}{2}S \nonumber \\
&& \hspace*{114pt} - \frac{4(3 Z_a^2 m_b^2 + Z_b^2 m_a^2)}{3 m_a m_b} L \bigg) \Bigg]\nonumber \\
& + \hspace*{-6pt} & {} \hspace*{-1pt} i 8\pi G  m_a m_b \hspace*{0.5pt} \alpha Z_a Z_b \hspace*{0.7pt} {L\over q^2} \sqrt{m_a m_b \over s-s_0} \hspace*{-2.5pt} \left(\hspace*{-1.4pt}\frac{3}{2} \hspace*{0.6pt}  {1\over m_a}\bar{u}(p_4 \hspace*{-0.2pt}) \hspace*{-1.2pt} \! \not \!{p}_1u(p_3\hspace*{-0.2pt}) \hspace*{-2.3pt} - \hspace*{-2pt} \frac{1}{2}\bar{u}(p_4\hspace*{-0.2pt})u(p_3\hspace*{-0.2pt})\hspace*{-3.8pt}\right) \hspace*{-3pt}. \nonumber \\ \quad \quad \quad \label{eq:su}
\end{eqnarray}
Defining the spin four-vector as
\begin{equation}
S_b^\mu={1\over 2}\bar{u}(p_4)\gamma_5\gamma^\mu u(p_3)
\end{equation}
where $\gamma_5=-i\gamma^0\gamma^1\gamma^2\gamma^3$ and taking
$\epsilon^{0123}=+1$, we find the identity
\begin{equation}
\bar{u}(p_4)\gamma_\mu u(p_3)=\left({1\over 1-{q^2\over
4m_b^2}}\right)\left[{(p_3+p_4)_\mu\over
2m_b}\bar{u}(p_4)u(p_3)-{i\over
m_b^2}\epsilon_{\mu\beta\gamma\delta}q^\beta p_3^\gamma
S_b^\delta\right]\label{eq:id}
\end{equation}
whereby the transition amplitude can be written in the form
\begin{eqnarray}
{}^{1\over 2}{\cal M}_{tot}^{(2)}(q)\hspace*{-5pt}&=&\hspace*{-5pt} G \alpha Z_a Z_b \Bigg[{\mathcal U}_b \Big(-6 (m_a+m_b)S + 12 L \Big) \nonumber \\
&& \hspace*{36pt} +  i \frac{{\mathcal E}_b}{m_a m_b^2} \hspace*{-2pt} \left( \hspace*{-1pt}\left( \hspace*{-1pt} -\frac{2 m_a m_b (2 m_a \hspace*{-1.5pt} + \hspace*{-1.5pt} 3 m_b)}{s - s_0} \hspace*{-1pt} -5 m_a \hspace*{-1pt} - \hspace*{-1pt} \frac{15}{2} m_b \hspace*{-2pt}\right) \hspace*{-2pt} S \right.\nonumber \\
&& \hspace*{88pt} + \frac{32}{3} L \hspace*{-1pt}\Bigg) \Bigg]\nonumber \\
& + & \hspace*{-5pt} G \alpha \Bigg[{\mathcal U}_b \left( - (Z_a^2 m_b + Z_b^2 m_a) S - \frac{8(Z_a^2 m_b^2 + Z_b^2 m_a^2)}{3 m_a m_b} \hspace*{1pt} L \right) \nonumber \\
 && \hspace*{10pt} + i \frac{{\mathcal E}_b}{m_a m_b^2} \bigg(\hspace*{-2pt} - \left(\frac{3}{2} Z_a^2 m_b + Z_b^2 m_a \hspace*{-1pt}\right) \hspace*{-1pt}S - \frac{4 (3 Z_a^2 m_b^2 + Z_b^2 m_a^2)}{3 m_a m_b} \hspace*{1pt} L \hspace*{-1pt}\bigg) \hspace*{-1pt} \Bigg] \nonumber \\
&+& \hspace*{-5pt} i 8\pi G  m_a m_b \, \alpha Z_a Z_b \hspace*{1.5pt} {L\over q^2} \sqrt{m_a m_b \over s-s_0} \left({\mathcal U}_b + \frac{3}{2} \, i \frac{{\mathcal E}_b}{m_a m_b^2}\right) \label{eq_mix_Mh_rel}
\end{eqnarray}
using the shorthand notation
\begin{equation}
{\mathcal U}_b \equiv \bar{u}(p_4)u(p_3)
\quad \quad \mbox{and} \quad \quad
{\mathcal E}_b \equiv \epsilon_{\alpha\beta\gamma\delta}p_1^\alpha
p_3^\beta q^\gamma S_b^\delta.
\end{equation}
This form has the advantage that one can easily read off the
spin-independent component of the amplitude (which is the same as
the spin-averaged amplitude) by setting ${\mathcal U}_b \rightarrow
1$ and ${\mathcal E}_b \rightarrow 0$ (see Eqs. (\ref{eq:nr}) and
(\ref{eq:epsireduce}) below), and we observe that the
spin-independent part of Eq. (\ref{eq_mix_Mh_rel}) and the amplitude
for spin-0 -- spin-0 scattering in Eq. (\ref{eq_amp_00}) have
identical forms.

Before we can define an appropriate second order potential, we must,
of course, remove the imaginary parts as done in the spinless
scattering case above.  However, we observe that when one or more
particles carries spin a new from appears, proportional to
$1/(s-s_0)$, which also must be removed before we can produce a
proper higher order potential.  Both forms must be eliminated by
subtraction of the iterated lowest order potential as before. Before
seeing how this is done, we first write the nonrelativistic
amplitude in the symmetric center of mass frame ($\vec p_1 = - \vec
p_3 = \vec p + \vec q /2$) where
\begin{equation}
S_b^\alpha\stackrel{NR}{\longrightarrow}(0,\vec{S}_b) \ \ \ \ \ {\rm
with} \ \ \ \ \ \vec{S}_b = {1\over
2}\chi_f^{b\dagger}\vec{\sigma}\chi_i^b,
\end{equation}
\begin{equation}
\bar{u}(p_4)u(p_3) \stackrel{NR}{\longrightarrow}
\chi_f^{b\dagger}\chi_i^b-{i\over 2m_b^2}\vec{S}_b\cdot \vec{p}
\times\vec{q} , \label{eq:nr}
\end{equation}
\begin{equation} \label{eq:epsireduce}
\epsilon_{\alpha\beta\gamma\delta}p_1^\alpha p_3^\beta q^\gamma
S_b^\delta\stackrel{NR}{\longrightarrow}(m_a+m_b)\left(1+\frac{\vec
p^{\hspace*{1.4pt}2}}{2 m_a
m_b}\right)\vec{S}_b\cdot\vec{p}\times\vec{q}
\end{equation}
and
\begin{equation}
 \frac{1}{s-s_0} \stackrel{NR}{\longrightarrow} \frac{m_a m_b}{(m_a + m_b)^2} \frac{1}{p_0^2} + \frac{(m_a - m_b)^2}{4 m_a m_b (m_a + m_b)^2}.
\end{equation}
We find then
\begin{eqnarray}
{}^{1\over 2}{\cal M}_{tot}^{(2)}(\vec q)\hspace*{-7pt} &\simeq\hspace*{-7pt}&\Bigg[ G \alpha Z_a Z_b
\Big(\hspace*{-3pt} - \hspace*{-1pt} 6 (m_a \hspace*{-1pt} + \hspace*{-1pt} m_b)S \hspace*{-1pt} + \hspace*{-1pt} 12 L \Big) \hspace*{-1pt} + i 8\pi G  m_a m_b \, \alpha Z_a Z_b \hspace*{1.5pt} {L\over q^2} \frac{m_r}{p_0} \nonumber \\
&& + \hspace*{1pt} G \alpha \left( - (Z_a^2 m_b + Z_b^2 m_a) S - \frac{8(Z_a^2 m_b^2 + Z_b^2 m_a^2)}{3 m_a m_b} \hspace*{1pt} L \right)\Bigg]
\chi_f^{b\dagger}\chi_i^b \nonumber\\
&+\hspace*{-7pt}&\Bigg[\hspace*{-0.8pt} \frac{G \alpha Z_a Z_b}{m_a m_b} \hspace*{-2.5pt} \left(\hspace*{-2.6pt} - \frac{3m_a^3 \hspace*{-2.2pt} + \hspace*{-2.4pt} 13 m_a^2 m_b \hspace*{-2.2pt} + \hspace*{-2.5pt} 18 m_a m_b^2
 \hspace*{-2.1pt} + \hspace*{-2.1pt} 9 m_b^3}{m_a \hspace*{-1pt} + \hspace*{-1pt} m_b}  \hspace*{0.5pt} S \hspace*{-1pt}
+ \hspace*{-1.3pt} \frac{14 m_a \hspace*{-2pt} + \hspace*{-2pt} 32 m_b}{3}  \hspace*{0.5pt} L \hspace*{-2.2pt} \right)\nonumber\\
&&- \frac{2 G m_a m_b \hspace*{1pt} \alpha Z_a Z_b (2 m_a + 3 m_b)}{(m_a + m_b)} \left(- i \frac{2 \pi
L}{p_0 q^2} + \frac{S}{p_0^2} \right) \nonumber \\
&& + \frac{G \alpha}{m_a m_b} \bigg( \hspace*{-3.5pt} - \hspace*{-1pt} \frac{1}{2}\Big(Z_a^2 (2 m_a m_b + 3 m_b^2) + Z_b^2 (m_a^2 + 2 m_a m_b)\Big)S \nonumber \\
&& \hspace*{49pt} - \frac{4(Z_a^2(2 m_a m_b + 3 m_b^2) + Z_b^2 m_a^2)}{3 m_a} L \bigg)\Bigg] {i\over
m_b}\vec{S}_b\cdot\vec{p}\times\vec{q}. \nonumber \\ \quad \label{eq_mix_Mh_nr}
\end{eqnarray}
Again, the spin-independent component---the piece proportional to
$\chi_f^{b\dagger}\chi_i^b$---agrees with the corresponding
expression for spinless scattering in Eq. (\ref{eq_amp_00_nr}). The
spin-dependent component is new and has the form of a spin-orbit
coupling.

In order to remove the imaginary---$i L /(q^2 p_0)$---and real---$S
/ p_0^2$---pieces, we iterate the lowest order potential, as done in
the spinless case, but we note that in the case of a system with
spin these lowest order potentials have, in addition to the usual
spin-independent Newton/Coulomb pieces, a spin-dependent component
which is of spin-orbit character.  We refer the reader to
\cite{hrem} and \cite{hrgr} for the explicit derivation of the
leading order potentials for spin-0 -- spin-1/2 scattering. These
lowest order potentials split into spin-independent ($S-I$) and
spin-dependent ($S-O$) components---
\begin{align}
\left<\vec p_f \left| {}^{\frac{1}{2}} \hat V^{(1)}_C \right| \vec
p_i \, \right> & =
 \left<\vec p_f \left| {}^{\frac{1}{2}} \hat V^{(1)}_{C, \hspace*{1pt} S-I} \right| \vec p_i \, \right>
 + \left<\vec p_f \left| {}^{\frac{1}{2}} \hat V^{(1)}_{C, \hspace*{1pt} S-O} \right| \vec p_i \, \right> \notag \\
\left<\vec p_f \left| {}^{\frac{1}{2}} \hat V^{(1)}_G \right| \vec
p_i \, \right> & =
 \left<\vec p_f \left| {}^{\frac{1}{2}} \hat V^{(1)}_{G, \hspace*{1pt} S-I} \right| \vec p_i \, \right>
 + \left<\vec p_f \left| {}^{\frac{1}{2}} \hat V^{(1)}_{G, \hspace*{1pt} S-O} \right| \vec p_i \, \right>
\end{align}
where
\begin{eqnarray}
 \left<\vec p_f \left| {}^{\frac{1}{2}} \hat V^{(1)}_{C, \hspace*{1pt} S-I} \right| \vec p_i \, \right> \hspace*{-5pt}
 &=\hspace*{-5pt}&{c_C^2 \over \vec{q}^{\hspace*{1.4pt}2}} \, \chi_f^{b\dagger}\chi_i^b
  = \frac{c_C^2}{(\vec p_i - \vec p_f)^2} \, \chi_f^{b\dagger}\chi_i^b \nonumber\\
 \left<\vec p_f \left| {}^{\frac{1}{2}} \hat V^{(1)}_{C, \hspace*{1pt} S-O} \right| \vec p_i \, \right> \hspace*{-5pt}
 &=\hspace*{-5pt}&{c_C^2 \over \vec{q}^{\hspace*{1.4pt}2}}{m_a+2m_b\over 2m_am_b}\, \frac{i}{m_b}\vec{S}_b\cdot\vec{p}\times\vec{q} \nonumber\\
 &=\hspace*{-5pt}&{c_C^2 \over (\vec p_i - \vec p_f)^2}{m_a+2m_b\over 2m_am_b}\, \frac{i}{m_b}\vec{S}_b\cdot \frac{1}{2} (\vec p_i + \vec p_f) \times (\vec p_i - \vec p_f) \nonumber\\
 \left<\vec p_f \left| {}^{\frac{1}{2}} \hat V^{(1)}_{G, \hspace*{1pt} S-I} \right| \vec p_i \, \right> \hspace*{-5pt}
 &=\hspace*{-5pt}& {c_G^2 \over \vec{q}^{\hspace*{1.4pt}2}} \, \chi_f^{b\dagger}\chi_i^b
  = \frac{c_G^2}{(\vec p_i - \vec p_f)^2} \, \chi_f^{b\dagger}\chi_i^b \nonumber\\
 \left<\vec p_f \left| {}^{\frac{1}{2}} \hat V^{(1)}_{G, \hspace*{1pt} S-O} \right| \vec p_i \, \right> \hspace*{-5pt}
 &=\hspace*{-5pt}& {c_G^2 \over \vec{q}^{\hspace*{1.4pt}2}}{3 m_a+ 4 m_b\over 2m_am_b}\, \frac{i}{m_b}\vec{S}_b\cdot\vec{p}\times\vec{q} \nonumber\\
 &=\hspace*{-5pt}& {c_G^2 \over (\vec p_i - \vec p_f)^2}{3 m_a + 4 m_b\over 2m_am_b}\, \frac{i}{m_b}\vec{S}_b\cdot \frac{1}{2} (\vec p_i + \vec p_f) \times (\vec p_i - \vec p_f) \nonumber\\
\end{eqnarray}
with $c_C^2 \equiv 4 \pi \alpha Z_a Z_b$ and $c_G^2 \equiv - 4 \pi G
m_a m_b$. We find then that the iterated amplitude splits also into
spin-independent and spin-dependent pieces. The leading
spin-independent iteration amplitude is
\begin{eqnarray}
{}^{1\over 2}{\rm Amp}^{(2)}_{S-I}(\vec q) \hspace*{-9pt}{} &=\hspace*{-5pt}&-
\int{d^3\ell\over (2\pi)^3} \, \frac{\left<\vec p_f \left|
{}^{\frac{1}{2}} \hat V^{(1)}_{G, \hspace*{1pt} S-I} \right| \vec \ell \, \right>
\left<\vec \ell \left| {}^{\frac{1}{2}}
 \hat V^{(1)}_{C, \hspace*{1pt} S-I} \right| \vec p_i \right>}{\frac{p_0^2}{2 m_r} -
 \frac{\ell^2}{2 m_r} + i \epsilon} \nonumber\\
&& - \int{d^3\ell\over (2\pi)^3} \, \frac{\left<\vec p_f \left|
{}^{\frac{1}{2}} \hat V^{(1)}_{C, \hspace*{1pt} S-I} \right| \vec \ell \, \right>
\left<\vec \ell \left| {}^{\frac{1}{2}}
 \hat V^{(1)}_{G, \hspace*{1pt} S-I} \right| \vec p_i \right>}{\frac{p_0^2}{2 m_r} -
 \frac{\ell^2}{2 m_r} + i \epsilon} \nonumber\\
&=\hspace*{-5pt}&i \sum_{s_\ell} \hspace*{-1pt} \int \hspace*{-1pt} {d^3\ell\over (2\pi)^3}
{c_G^2 \, \chi_f^{b\dagger}\chi_{s_\ell}^b \over |\vec{\ell} - \vec{p}_f|^2 + \lambda^2}
{i\over {p_0^2\over 2m_r}-{\ell^2\over 2m_r}+i\epsilon} {c_C^2 \, \chi_{s_\ell}^{b\dagger}\chi_i^b \over |\vec{p}_i- \vec{\ell} \hspace*{1pt}|^2+\lambda^2}\nonumber\\
&+\hspace*{-5pt}&i \sum_{s_\ell} \hspace*{-1pt} \int \hspace*{-1pt} {d^3\ell\over (2\pi)^3}
{c_C^2 \, \chi_f^{b\dagger}\chi_{s_\ell}^b \over |\vec{\ell} - \vec{p}_f|^2 + \lambda^2}
{i\over {p_0^2\over 2m_r}-{\ell^2\over 2m_r}+i\epsilon} {c_G^2 \, \chi_{s_\ell}^{b\dagger}\chi_i^b \over |\vec{p}_i- \vec{\ell} \hspace*{1pt}|^2+\lambda^2}\nonumber\\
&\stackrel{\lambda\rightarrow 0} {\longrightarrow} \hspace*{-5pt} &\chi_f^{b\dagger}\chi_i^b \, 2 H
= i 8\pi G  m_a m_b \, \alpha Z_a Z_b \hspace*{1.5pt} {L\over q^2} \frac{m_r}{p_0} \chi_f^{b\dagger}\chi_i^b
\label{eq:iteration0hSI}
\end{eqnarray}
while the leading spin-dependent term is
\begin{eqnarray}
{}^{1\over 2}{\rm Amp}^{(2)}_{S-O}(\vec q) \hspace*{-10.3pt} &= \hspace*{-10pt} &-
\int{d^3\ell\over (2\pi)^3} \, \frac{\left<\vec p_f \left|
{}^{\frac{1}{2}} \hat V^{(1)}_{G, \hspace*{1pt} S-I}
 \right| \vec \ell \, \right> \left<\vec \ell \left| {}^{\frac{1}{2}}
 \hat V^{(1)}_{C, \hspace*{1pt} S-O} \right| \vec p_i \right>}{\frac{p_0^2}{2 m_r} -
 \frac{\ell^2}{2 m_r} + i \epsilon} \nonumber\\
&\hspace*{-5pt}&- \int{d^3\ell\over (2\pi)^3} \, \frac{\left<\vec p_f \left|
{}^{\frac{1}{2}} \hat V^{(1)}_{C, \hspace*{1pt} S-O} \right| \vec \ell \, \right>
\left<\vec \ell \left| {}^{\frac{1}{2}} \hat V^{(1)}_{G, \hspace*{1pt} S-I} \right|
\vec p_i \right>}{\frac{p_0^2}{2 m_r} -
\frac{\ell^2}{2 m_r} + i \epsilon} \nonumber\\
&\hspace*{-5pt}&- \int{d^3\ell\over (2\pi)^3} \, \frac{\left<\vec p_f \left|
{}^{\frac{1}{2}} \hat V^{(1)}_{G, \hspace*{1pt} S-O} \right| \vec \ell \, \right>
\left<\vec \ell \left| {}^{\frac{1}{2}} \hat V^{(1)}_{C, \hspace*{1pt} S-I} \right|
\vec p_i \right>}{\frac{p_0^2}{2 m_r} -
\frac{\ell^2}{2 m_r} + i \epsilon} \nonumber\\
&\hspace*{-5pt}&- \int{d^3\ell\over (2\pi)^3} \, \frac{\left<\vec p_f \left|
{}^{\frac{1}{2}} \hat V^{(1)}_{C, \hspace*{1pt} S-I} \right| \vec \ell \, \right>
\left<\vec \ell \left| {}^{\frac{1}{2}} \hat V^{(1)}_{G, \hspace*{1pt} S-O} \right|
\vec p_i \right>}{\frac{p_0^2}{2 m_r} -
\frac{\ell^2}{2 m_r} + i \epsilon} \nonumber\\
&=\hspace*{-10pt}&{i(m_a+2m_b)\over 2 m_a m_b^2}
\vec{S}_b \cdot \nonumber\\
&\hspace*{-9pt}& \left(\hspace*{-2pt}i \hspace*{-3.2pt} \int \hspace*{-3.2pt}
{d^3\ell\over (2\pi)^3} {c_G^2\over |\vec{\ell} \hspace*{-1.1pt} -
\hspace*{-1.1pt} \vec{p}_f|^2 \hspace*{-1.1pt} +
\hspace*{-1.2pt} \lambda^2}{i \over {p_0^2\over 2m_r}
\hspace*{-1.1pt} - \hspace*{-1.1pt} {\ell^2\over 2m_r}
\hspace*{-1.1pt} + \hspace*{-1.1pt} i\epsilon}{c_C^2 \,
\frac{1}{2}(\vec p_i \hspace*{-1.1pt} + \hspace*{-1.1pt} \vec \ell)
\hspace*{-2.5pt} \times \hspace*{-2.5pt} (\vec p_i \hspace*{-1.1pt}
- \hspace*{-1.1pt} \vec \ell)\over |\vec{p}_i \hspace*{-1.1pt} -
\hspace*{-1.1pt} \vec{\ell} \hspace*{1pt}|^2 \hspace*{-1.1pt}
 + \hspace*{-1.2pt} \lambda^2}\right.\nonumber\\
&\hspace*{-9pt}&\left. \hspace*{-3.6pt}+ i \hspace*{-3.2pt} \int \hspace*{-3.2pt}
{d^3\ell\over (2\pi)^3} {c_C^2 \, \frac{1}{2}(\vec \ell
\hspace*{-1.3pt} + \hspace*{-1.4pt} \vec p_f) \hspace*{-2.7pt}
\times \hspace*{-2.7pt} (\vec \ell \hspace*{-1.3pt} -
\hspace*{-1.4pt} \vec p_f)\over |\vec{\ell} \hspace*{-1.1pt} -
\hspace*{-1.1pt} \vec{p}_f|^2 \hspace*{-1.1pt} +
\hspace*{-1.2pt} \lambda^2}{i \over {p_0^2\over 2m_r}
\hspace*{-1.1pt} - \hspace*{-1.1pt} {\ell^2\over 2m_r}
\hspace*{-1.1pt} + \hspace*{-1.1pt} i\epsilon}{c_G^2\over |\vec{p}_i
\hspace*{-1.1pt} - \hspace*{-1.1pt} \vec{\ell} \hspace*{1pt}|^2 \hspace*{-1.1pt}
 + \hspace*{-1.2pt}\lambda^2}\hspace*{-3pt}\right)\nonumber\\
&+\hspace*{-10pt}&{i(3m_a+4m_b)\over 2 m_a m_b^2}
\vec{S}_b \cdot \nonumber\\
&\hspace*{-9pt}& \left(\hspace*{-2pt}i \hspace*{-3.2pt} \int \hspace*{-3.2pt}
{d^3\ell\over (2\pi)^3} {c_C^2\over |\vec{\ell} \hspace*{-1.1pt} -
\hspace*{-1.1pt} \vec{p}_f |^2 \hspace*{-1.1pt} +
\hspace*{-1.2pt} \lambda^2}{i \over {p_0^2\over 2m_r}
\hspace*{-1.1pt} - \hspace*{-1.1pt} {\ell^2\over 2m_r}
\hspace*{-1.1pt} + \hspace*{-1.1pt} i\epsilon}{c_G^2 \,
\frac{1}{2}(\vec p_i \hspace*{-1.1pt} + \hspace*{-1.1pt} \vec \ell)
\hspace*{-2.5pt} \times \hspace*{-2.5pt} (\vec p_i \hspace*{-1.1pt}
- \hspace*{-1.1pt} \vec \ell)\over |\vec{p}_i \hspace*{-1.1pt} -
\hspace*{-1.1pt} \vec{\ell} \hspace*{1pt}|^2 \hspace*{-1.1pt}
 + \hspace*{-1.2pt} \lambda^2}\right.\nonumber\\
&\hspace*{-9pt}&\left. \hspace*{-3.6pt}+ i \hspace*{-3.2pt} \int \hspace*{-3.2pt}
{d^3\ell\over (2\pi)^3} {c_G^2 \, \frac{1}{2}(\vec \ell
\hspace*{-1.3pt} + \hspace*{-1.4pt} \vec p_f) \hspace*{-2.7pt}
\times \hspace*{-2.7pt} (\vec \ell \hspace*{-1.3pt} -
\hspace*{-1.4pt} \vec p_f)\over |\vec{\ell} \hspace*{-1.1pt} -
\hspace*{-1.1pt} \vec{p}_f|^2 \hspace*{-1.1pt} +
\hspace*{-1.2pt} \lambda^2}{i \over {p_0^2\over 2m_r}
\hspace*{-1.1pt} - \hspace*{-1.1pt} {\ell^2\over 2m_r}
\hspace*{-1.1pt} + \hspace*{-1.1pt} i\epsilon}{c_C^2\over |\vec{p}_i
\hspace*{-1.1pt} - \hspace*{-1.1pt} \vec{\ell} \hspace*{1pt}|^2 \hspace*{-1.1pt}
 + \hspace*{-1.2pt}\lambda^2}\hspace*{-3pt}\right)\nonumber\\
&\stackrel{\lambda\rightarrow 0}{\longrightarrow}\hspace*{-8pt}&{}\, {i(4 m_a+6m_b)\over
2 m_a m_b^2} \, \vec{S}_b\cdot\vec{H}\times
\vec{q}\nonumber\\
&=\hspace*{-10pt}& - \frac{2 G m_a m_b \hspace*{1pt} \alpha Z_a Z_b (2 m_a \hspace*{-1pt} + \hspace*{-1pt} 3 m_b)}{(m_a + m_b)} \hspace*{-1pt} \left( \hspace*{-2pt} - i \frac{2 \pi
L}{p_0 q^2} \hspace*{-1pt} + \hspace*{-1pt} \frac{S}{p_0^2} \hspace*{-1pt} \right) {i\over
m_b}\vec{S}_b\cdot\vec{p}\times\vec{q} \nonumber \\ \quad \label{eq:iteration0hSO}
\end{eqnarray}
(In principle one has to also include the iteration with two
spin-orbit components of the leading potentials.  However, this
procedure but yields terms of higher order in $p_0^2$.)

We observe then that the second order Born term for spin-0 -- spin-1/2
scattering is precisely of the form required to remove the offending
imaginary $i L /(q^2 p_0)$ {\it and} real $S / p_0^2$ pieces from
the second order scattering amplitude in Eq. (\ref{eq_mix_Mh_nr}).
What remains can be Fourier-transformed to yield a well-defined
second order potential of the form
\begin{eqnarray}
{}^{\frac{1}{2}}V_{CG}^{(2)}(\vec r) \hspace*{-5pt} &=&-\int{d^3q\over (2\pi)^3}e^{-i\vec{q}\cdot\vec{r}}
\left[{}^{1\over 2}{\cal M}^{(2)}_{tot}(\vec{q})-{}^{1\over 2}{\rm Amp}_{CG}^{(2)}(\vec{q}) \right]\nonumber\\
& \simeq \hspace*{-5pt} & G \alpha \Bigg[\frac{1}{2} \frac{ (Z_a^2 m_b + Z_b^2 m_a)}{r^2} + 3 \frac{Z_a Z_b (m_a + m_b)}{r^2} \nonumber \\
&& {} \hspace*{15pt} - \frac{4 \hbar}{{3 \pi r^3}} \left(Z_b^2 \frac{m_a}{m_b} + Z_a^2 \frac{m_b}{m_a}\right) + \frac{6 Z_a Z_b \hbar}{\pi r^3} \Bigg] \chi_f^{b\dagger}\chi_i^b \nonumber \\
&+\hspace*{-5pt} & \Bigg[- \frac{Z_a Z_b G \alpha(3 m_a^3 + 13 m_a^2 m_b + 18 m_a m_b^2 + 9 m_b^3)}{m_a m_b^2 (m_a + m_b) \hspace*{1pt} r^4}  \nonumber \\
&& {}- \frac{G \alpha(Z_a^2(2 m_a m_b + 3 m_b^2) + Z_b^2 (m_a^2 + 2 m_a m_b))}{2 m_a m_b^2 \hspace*{1pt} r^4} \nonumber \\
&& {} - \frac{Z_a Z_b G \alpha \hbar (7 m_a \hspace*{-2pt} + \hspace*{-2pt} 16 m_b)}{\pi m_a m_b^2 r^5} \nonumber \\
&& {} + \hspace*{-1pt} \frac{2 G \alpha \hbar (Z_a^2(2 m_a m_b \hspace*{-2pt} + \hspace*{-2pt} 3 m_b^2) \hspace*{-2pt} + \hspace*{-2pt} Z_b^2 m_a^2)}{\pi m_a^2 m_b^2 r^5} \hspace*{-1pt}\Bigg] \hspace*{-1pt} \vec L \hspace*{-1.5pt} \cdot \hspace*{-1.5pt} \vec S_b.
 \quad \label{eq:hd}
\end{eqnarray}
where $\vec L \equiv \vec r \times \vec p$ is the angular momentum
and $\vec r \equiv \vec r_a - \vec r_b$. Comparing with the second
order potential found in the spinless case---Eq. (\ref{eq:so})---we
observe that the {\it spin-independent} component of the potential
${}^{1\over 2}V_{CG}^{(2)}(\vec{r})$ ({\it i.e.}, the piece
proportional to $\chi_f^{b\dagger}\chi_i^b$) is {\it identical} to
the potential ${}^0V_{CG}^{(2)}(\vec{r})$ found in the case of
spinless scattering.  This is the universality property that we had
expected and verifies the correctness of our calculation.  However,
there also exists now a spin-orbit component of the potential having
a new kinematic form, which we suspect is also universal. In order
to check this assumption, we evaluate the case of a spin-1/2
particle $a$ scattering from a spin-1/2 particle $b$, the
spin-independent piece of which was calculated by Butt \cite{msb}.
Before leaving this section we note that all results were deriving
using a g-factor of $g=2$ at tree level throughout for the photon
couplings, which is the natural value for spin-1/2 particle arising
from the Dirac Lagrangian \cite{Holstein:2006wi}.  In particular, we
observe that the form of the spin-orbit potential does depend on
this choice. See \cite{hrem} for a more detailed discussion on the
$g$-dependence of spin-dependent components in the case of purely
electromagnetic scattering.

\section{Spin-1/2 -- Spin-1/2 Scattering}

As our next calculation, we evaluate the scattering of a pair of
spin-1/2 particles.  The spin-averaged version of this calculation
was performed by Butt and was claimed not to obey universality \cite{msb}. This
is one reason that we wish to examine this system in detail.

As before we list the result of the calculation and refer to the
appendices for the Feynman rules and the relevant loop integrals.
Using the spin-1/2 identity Eq. (\ref{eq:id}) and its pendant for
particle $a$
\begin{equation}
\bar{u}(p_2)\gamma_\mu u(p_1)=\left({1\over 1-{q^2\over
4m_a^2}}\right)\left[{(p_1+p_2)_\mu\over
2m_a}\bar{u}(p_2)u(p_1)+{i\over
m_a^2}\epsilon_{\mu\beta\gamma\delta}q^\beta p_1^\gamma
S_a^\delta\right]\label{eq:ida}
\end{equation}
where we have defined the spin vector
$$S_{a}^\mu={1\over 2}\bar{u}(p_2)\gamma_5\gamma^\mu u(p_1)$$
we present our results, as before, diagram by diagram:
\begin{eqnarray}
{}^{{1\over 2}{1\over 2}}{\cal M}_{\ref{fig_diags}a}^{(2)}(q)\hspace*{-5pt}&=&\hspace*{-5pt} 0 \nonumber \\
{}^{{1\over 2}{1\over 2}}{\cal M}_{\ref{fig_diags}b}^{(2)}(q)\hspace*{-5pt}&=&\hspace*{-5pt} G \alpha Z_a Z_b \Bigg[{\mathcal U}_a {\mathcal U}_b \left(-3 L - 6 m_a S\right) \nonumber\\
 && \hspace*{37pt} +  i \frac{{\mathcal E}_a {\mathcal U}_b}{m_a^2 m_b} \left( L - 2m_a S\right)
 +  i \frac{{\mathcal U}_a {\mathcal E}_b}{m_a m_b^2} \left(-3 L - 6 m_a S\right) \nonumber \\
 && \hspace*{37pt} + \frac{S_a\cdot qS_b\cdot q}{m_a m_b} \left(L - 2 m_a S\right) - \frac{q^2 S_a \cdot S_b}{m_a m_b} \left(L - 2 m_a S\right) \hspace*{-1.5pt}\Bigg]\nonumber\\
{}^{{1\over 2}{1\over 2}}{\cal M}_{\ref{fig_diags}c}^{(2)}(q)\hspace*{-5pt}&=&\hspace*{-5pt} G \alpha Z_a Z_b \Bigg[{\mathcal U}_a {\mathcal U}_b \left(-3 L - 6 m_b S\right) \nonumber\\
 && \hspace*{37pt} +  i \frac{{\mathcal E}_a {\mathcal U}_b}{m_a^2 m_b} \left(-3 L - 6m_b S\right)
 +  i \frac{{\mathcal U}_a {\mathcal E}_b}{m_a m_b^2} \left(L - 2 m_b S\right) \nonumber \\
 && \hspace*{37pt} + \frac{S_a\cdot qS_b\cdot q}{m_a m_b} \left(L - 2 m_b S\right)
    - \frac{q^2 S_a \cdot S_b}{m_a m_b} \left(L - 2 m_b S\right) \hspace*{-1.5pt}\Bigg]\nonumber\\
{}^{{1\over 2}{1\over 2}}{\cal M}_{\ref{fig_diags}d}^{(2)}(q)\hspace*{-5pt}&=&\hspace*{-5pt} G \alpha Z_a Z_b \Bigg[{\mathcal U}_a {\mathcal U}_b
\hspace*{-1.5pt}\Bigg(\hspace*{-1.5pt}L \hspace*{-1pt}\left(\hspace*{-2.5pt} - {8 m_a m_b\over q^2}\hspace*{-1pt}-\hspace*{-1pt}\frac{21 m_a^2\hspace*{-1pt} + 3\hspace*{-1pt} m_a m_b\hspace*{-1pt} + \hspace*{-1pt} 21 m_b^2}{3 m_am_b}\hspace*{-0.5pt}\right) \nonumber \\
 && \hspace*{68pt} + S \left(-6 (m_a + m_b)\right)\hspace*{-1pt} \Bigg) \nonumber\\
&&\hspace*{37pt}+  i \frac{{\mathcal E}_a {\mathcal U}_b}{m_a^2 m_b}\Bigg(\hspace*{-1pt} L \hspace*{-1pt} \left(\hspace*{-2.5pt} - \frac{12 m_a m_b}{q^2} \hspace*{-1pt}
  - \hspace*{-1pt} \frac{27 m_a^2 + 4 m_a m_b + 31 m_b^2}{3 m_a m_b}\right) \nonumber\\
&& \hspace*{82pt} + S \hspace*{-1pt} \left( \hspace*{-2.5pt} - \frac{2 m_a m_b(3 m_a \hspace*{-2pt} + \hspace*{-1.5pt} 2 m_b)}{s-s_0} \hspace*{-1pt} - \hspace*{-1pt} (10 m_a \hspace*{-2pt} + \hspace*{-1.5pt} 7 m_b) \hspace*{-1.8pt}\right) \! \hspace*{-1.5pt} \hspace*{-2pt} \Bigg)\nonumber\\
&&\hspace*{37pt}+  i \frac{{\mathcal U}_a {\mathcal E}_b}{m_a m_b^2}\Bigg(\hspace*{-1pt} L \hspace*{-1pt} \left(\hspace*{-2.5pt} - \frac{12 m_a m_b}{q^2}\hspace*{-1pt} - \hspace*{-1pt} \frac{31 m_a^2 - 4 m_a m_b + 27 m_b^2}{3 m_a m_b}\right) \nonumber\\
&& \hspace*{82pt} +S \hspace*{-1pt} \left(\hspace*{-2.5pt} - \frac{2 m_a m_b(2 m_a \hspace*{-2pt} + \hspace*{-1.5pt} 3 m_b)}{s-s_0} \hspace*{-1pt} - \hspace*{-1pt} (7 m_a \hspace*{-2pt} + \hspace*{-1.5pt} 10 m_b) \hspace*{-1.8pt}\right) \! \hspace*{-1.5pt} \hspace*{-2pt} \Bigg)\nonumber\\
&&\hspace*{37pt} + \frac{S_a \hspace*{-2pt} \cdot \hspace*{-1.5pt} qS_b \hspace*{-2pt} \cdot \hspace*{-1.5pt} q}{m_a m_b} \Bigg( \hspace*{-1.7pt} L \hspace*{-1.5pt} \left(\hspace*{-2.5pt} - \frac{8 m_a m_b}{q^2} \hspace*{-1pt} - \hspace*{-1pt} \frac{20m_a^2 \hspace*{-1.5pt} + \hspace*{-1.5pt} 18 m_a m_b \hspace*{-1.5pt} + \hspace*{-1.5pt} 20 m_b^2}{3 m_a m_b} \hspace*{-1pt} \right) \nonumber\\
&& \hspace*{96pt} + S \hspace*{-1pt} \left(\hspace*{-2.5pt} - \frac{2 m_a m_b (m_a \hspace*{-1.7pt} + \hspace*{-1.7pt} m_b)}{s - s_0} \hspace*{-1pt} - \hspace*{-1pt} \frac{43}{4} \hspace*{-1pt} \left(m_a \hspace*{-1.7pt} + \hspace*{-1.7pt} m_b \right) \hspace*{-1.5pt} \right) \! \hspace*{-3.5pt} \Bigg) \nonumber \\
&&\hspace*{37pt} - \frac{q^2 S_a \cdot S_b}{m_a m_b} \Bigg( \hspace*{-1.5pt} L \hspace*{-2pt} \left(\hspace*{-2.5pt} - \frac{4 m_a m_b}{q^2} \hspace*{-1.5pt} - \hspace*{-1.5pt} \frac{37 m_a^2 \hspace*{-1.2pt} + \hspace*{-1.2pt} 62 m_a m_b \hspace*{-1.2pt} + \hspace*{-1.2pt} 37 m_b^2}{6 m_a m_b} \hspace*{-1pt}\right)\nonumber\\
&& \hspace*{92pt} + S \hspace*{-1pt} \left( \hspace*{-1pt} - \frac{2 m_a m_b (m_a \hspace*{-1.7pt} + \hspace*{-1.7pt} m_b)}{s - s_0} \hspace*{-1pt} - \hspace*{-1pt} \frac{41}{4} \left(m_a \hspace*{-1.7pt} + \hspace*{-1.7pt} m_b \right) \hspace*{-1.5pt} \right) \hspace*{-3.5pt} \! \Bigg) \nonumber \\
&&\hspace*{37pt} + \Big(2 S_a \hspace*{-0.5pt} \cdot \hspace*{-0.5pt} p_3 S_b \hspace*{-0.5pt} \cdot \hspace*{-0.5pt} p_1 \hspace*{-0.5pt} + \hspace*{-0.5pt} S_a \hspace*{-0.5pt} \cdot \hspace*{-0.5pt} q S_b \hspace*{-0.5pt} \cdot \hspace*{-0.5pt} p_1 \hspace*{-0.5pt} - \hspace*{-0.5pt} S_a \hspace*{-0.5pt} \cdot \hspace*{-0.5pt} p_3 S_b \hspace*{-0.5pt} \cdot \hspace*{-0.5pt} q\Big) \frac{- 26 L}{3 m_a m_b}\Bigg] \nonumber\\
&+& \hspace*{-5pt} i 8\pi G  m_a m_b \, \alpha Z_a Z_b \hspace*{1.5pt} {L\over q^2} \sqrt{m_a m_b \over s-s_0} \,
\Bigg({\mathcal U}_a {\mathcal U}_b + \frac{3}{2} \hspace*{1pt} i \frac{{\mathcal E}_a {\mathcal U}_b}{m_a^2 m_b} + \frac{3}{2} \hspace*{1pt} i \frac{{\mathcal U}_a {\mathcal E}_b}{m_a m_b^2}\nonumber\\
&&\hspace*{153pt}  + \frac{S_a \cdot qS_b \cdot q - \frac{1}{2}q^2 S_a \cdot S_b}{m_a m_b}\Bigg) \nonumber\\
{}^{{1\over 2}{1\over 2}}{\cal M}_{\ref{fig_diags}e}^{(2)}(q)\hspace*{-5pt}&=&\hspace*{-5pt} G \alpha Z_a Z_b \Bigg[{\mathcal U}_a {\mathcal U}_b
\hspace*{-1.5pt}\Bigg(\hspace*{-1.5pt}L \hspace*{-1pt}\left(\hspace*{-1pt} {8 m_a m_b\over q^2}\hspace*{-1pt}+\hspace*{-1pt}\frac{21 m_a^2\hspace*{-1pt} + \hspace*{-1pt} 53 m_a m_b\hspace*{-1pt} + \hspace*{-1pt} 21 m_b^2}{3 m_am_b}\hspace*{-0.5pt}\right) \nonumber \\
 && \hspace*{68pt} + S \ 4 (m_a + m_b)\hspace*{-1pt} \Bigg) \nonumber\\
&&\hspace*{37pt}+  i \frac{{\mathcal E}_a {\mathcal U}_b}{m_a^2 m_b}\Bigg(\hspace*{-1pt} L \hspace*{-1pt} \left(\hspace*{-1pt} \frac{12 m_a m_b}{q^2} \hspace*{-1pt}
  + \hspace*{-1pt} \frac{27 m_a^2 + 38 m_a m_b + 31 m_b^2}{3 m_a m_b}\right) \nonumber\\
&& \hspace*{82pt} + S \hspace*{-1pt} \left(\frac{7}{2} m_a \hspace*{-0.5pt} + \hspace*{-0.5pt} 6 m_b \right) \! \hspace*{-2pt} \Bigg)\nonumber\\
&&\hspace*{37pt}+  i \frac{{\mathcal U}_a {\mathcal E}_b}{m_a m_b^2}\Bigg(\hspace*{-1pt} L \hspace*{-1pt} \left(\hspace*{-1pt} \frac{12 m_a m_b}{q^2}\hspace*{-1pt} + \hspace*{-1pt} \frac{31 m_a^2 + 38 m_a m_b + 27 m_b^2}{3 m_a m_b}\right) \nonumber\\
&& \hspace*{82pt} + S \hspace*{-1pt} \left(6 m_a + \frac{7}{2} m_b) \right) \! \hspace*{-2pt} \Bigg)\nonumber\\
&&\hspace*{37pt} + \frac{S_a\cdot qS_b\cdot q}{m_a m_b} \Bigg( \hspace*{-1pt} L \hspace*{-1pt} \left(\hspace*{-1pt} \frac{8 m_a m_b}{q^2} \hspace*{-1pt} + \hspace*{-1pt} \frac{20m_a^2 \hspace*{-1pt} + \hspace*{-1pt} 12 m_a m_b \hspace*{-1pt} + \hspace*{-1pt} 20 m_b^2}{3 m_a m_b} \right) \nonumber\\
&& \hspace*{99pt} + S \hspace*{-1pt} \left(\hspace*{-1pt} \frac{9}{4} \left(m_a \hspace*{-1pt} + \hspace*{-1pt} m_b \right) \hspace*{-1pt} \right) \! \hspace*{-2pt} \Bigg) \nonumber \\
&&\hspace*{37pt} - \frac{q^2 S_a \cdot S_b}{m_a m_b} \Bigg( \hspace*{-1.5pt} L \hspace*{-2pt} \left(\hspace*{-1pt} \frac{4 m_a m_b}{q^2} \hspace*{-1pt} + \hspace*{-1pt} \frac{37 m_a^2 \hspace*{-1.2pt} + \hspace*{-1.2pt} 42 m_a m_b \hspace*{-1.2pt} + \hspace*{-1.2pt} 37 m_b^2}{6 m_a m_b} \hspace*{-1pt}\right)\nonumber\\
&& \hspace*{92pt} + S \hspace*{-1pt} \left(\frac{7}{4} \left(m_a \hspace*{-1pt} + \hspace*{-1pt} m_b \right) \hspace*{-1pt} \right) \hspace*{-3pt} \Bigg) \nonumber \\
&&\hspace*{37pt} + \Big(2 S_a \hspace*{-0.5pt} \cdot \hspace*{-0.5pt} p_3 S_b \hspace*{-0.5pt} \cdot \hspace*{-0.5pt} p_1 \hspace*{-0.5pt} + \hspace*{-0.5pt} S_a \hspace*{-0.5pt} \cdot \hspace*{-0.5pt} q S_b \hspace*{-0.5pt} \cdot \hspace*{-0.5pt} p_1 \hspace*{-0.5pt} - \hspace*{-0.5pt} S_a \hspace*{-0.5pt} \cdot \hspace*{-0.5pt} p_3 S_b \hspace*{-0.5pt} \cdot \hspace*{-0.5pt} q\Big) \frac{26 L}{3 m_a m_b}\Bigg] \nonumber\\
{}^{{1\over 2}{1\over 2}}{\cal M}_{\ref{fig_diags}f_\gamma}^{(2)}(q)\hspace*{-5pt}&=&\hspace*{-5pt} G \alpha Z_a Z_b \Bigg[{\mathcal U}_a {\mathcal U}_b \Big(2 (m_a + m_b) S\Big) \nonumber\\
 && \hspace*{37pt} +  i \frac{{\mathcal E}_a {\mathcal U}_b}{m_a^2 m_b} \Big(\left( m_a + 2 m_b \right) \! S \Big)
  +  i \frac{{\mathcal U}_a {\mathcal E}_b}{m_a m_b^2} \Big(\left(2 m_a + m_b \right) \! S \Big) \nonumber \\
 && \hspace*{37pt} + \frac{S_a\cdot qS_b\cdot q}{m_a m_b} \left(m_a + m_b\right) S
  - \frac{q^2 S_a \cdot S_b}{m_a m_b} \left(m_a + m_b\right) S  \Bigg]\nonumber\\
{}^{{1\over 2}{1\over 2}}{\cal M}_{\ref{fig_diags}f_g}^{(2)}(q)\hspace*{-5pt}&=&\hspace*{-5pt} G \alpha \Bigg[{\mathcal U}_a {\mathcal U}_b \left( \hspace*{-2pt} -\frac{8(Z_a^2 m_b^2 + Z_b^2 m_a^2)}{3 m_a m_b} \hspace*{1pt} L - \left(Z_a^2 m_b + Z_b^2 m_a\right) S \right)\nonumber \\
 && \hspace*{10pt} +  i \frac{{\mathcal E}_a {\mathcal U}_b}{m_a^2 m_b} \bigg(\hspace*{-3.5pt} - \frac{4 (Z_a^2 m_b^2 + 3 Z_b^2 m_a^2)}{3 m_a m_b} - \left(Z_a^2 m_b + \frac{3}{2} Z_b^2 m_a \hspace*{-1pt}\right) \hspace*{-1.5pt}S \hspace*{-2pt}\bigg)\nonumber \\
 && \hspace*{10pt}+  i \frac{{\mathcal U}_a {\mathcal E}_b}{m_a m_b^2} \bigg(\hspace*{-3.5pt} - \frac{4 (3 Z_a^2 m_b^2 + Z_b^2 m_a^2)}{3 m_a m_b} - \left(\frac{3}{2} Z_a^2 m_b + Z_b^2 m_a \hspace*{-1pt}\right) \hspace*{-1.5pt}S \hspace*{-2pt}\bigg)\nonumber \\
 && \hspace*{10pt} + \frac{S_a \hspace*{-1pt} \cdot \hspace*{-1pt} q S_b \hspace*{-1pt} \cdot \hspace*{-1pt} q}{m_a m_b} \bigg(\hspace*{-4.5pt} - \hspace*{-1pt} \frac{2(Z_a^2 m_b^2 \hspace*{-1pt} + \hspace*{-1pt} Z_b^2 m_a^2)}{3 m_a m_b} \hspace*{1pt} L \hspace*{-1pt} - \hspace*{-1pt}  \frac{1}{2}\left(Z_a^2 m_b \hspace*{-1pt} + \hspace*{-1pt}Z_b^2 m_a\right) \hspace*{-1.5pt} S \hspace*{-2pt} \bigg) \nonumber \\
 && \hspace*{10pt} - \frac{q^2 S_a \cdot S_b}{m_a m_b} \bigg(\hspace*{-4.5pt} - \hspace*{-1pt} \frac{2(Z_a^2 m_b^2 \hspace*{-1pt} + \hspace*{-1pt} Z_b^2 m_a^2)}{3 m_a m_b} \hspace*{1pt} L \hspace*{-1pt} - \hspace*{-1pt}  \frac{1}{2}\left(Z_a^2 m_b \hspace*{-1pt} + \hspace*{-1pt}Z_b^2 m_a\right) \hspace*{-1.5pt} S \hspace*{-2pt} \bigg) \nonumber \\
{}^{{1\over 2}{1\over 2}}{\cal M}_{\ref{fig_diags}g}^{(2)}(q)\hspace*{-5pt}&=&\hspace*{-5pt} G \alpha Z_a Z_b \Bigg[{\mathcal U}_a {\mathcal U}_b  \hspace*{-1pt} \left(\hspace*{-1pt} \frac{4}{3}L \hspace*{-0.5pt} \right)  \hspace*{-2pt}
 +  i \frac{{\mathcal E}_a {\mathcal U}_b}{m_a^2 m_b} \hspace*{-1pt} \left(\hspace*{-1pt} \frac{4}{3} L \hspace*{-0.5pt} \right) \hspace*{-2pt}
 +  i \frac{{\mathcal U}_a {\mathcal E}_b}{m_a m_b^2} \hspace*{-1pt} \left(\hspace*{-1pt} \frac{4}{3} L \hspace*{-0.5pt} \right) \nonumber \\
 && \hspace*{37pt}+ \frac{S_a\cdot qS_b\cdot q}{m_a m_b} \left(\frac{4}{3} L\right)
 - \frac{q^2 S_a \cdot S_b}{m_a m_b} \left(\frac{4}{3} L\right) \hspace*{-1.5pt}\Bigg]
\end{eqnarray}
where we have defined
\begin{equation}
{\mathcal U}_a=\bar{u}(p_2)u(p_1) \qquad \qquad {\mathcal
U}_b=\bar{u}(p_4)u(p_3)
\end{equation}
and
\begin{equation}
{\mathcal E}_i = \epsilon_{\alpha\beta\gamma\delta}p_1^\alpha
p_3^\beta q^\gamma S_i^\delta
\end{equation}
with $i = a, b$ to keep our notation compact.

At this point we can compare with Butt's spin-averaged result \cite{msb} by
examining the spin-independent terms, which can be read off easily
from our results by taking the limit ${\mathcal U}_{a,b} \rightarrow
1$, ${\mathcal E}_{a,b} \rightarrow 0$ and $S_a\cdot qS_b\cdot q,
S_a \cdot S_b \rightarrow 0$. We do not concur with many of Butt's
diagrams and do not understand the reason for this disagreement.
Undaunted, we sum the results of the above individual contributions
and determine the total amplitude
\begin{eqnarray}
{}^{\frac{1}{2}\frac{1}{2}}{\cal M}_{tot}^{(2)}(q)\hspace*{-8.5pt}&=&\hspace*{-8pt} G \alpha Z_a Z_b \Bigg[{\mathcal U}_a {\mathcal U}_b \Big(-6 (m_a+m_b)S + 12 L \Big) \nonumber \\
&& \hspace*{32pt} +  i \frac{{\mathcal E}_a {\mathcal U}_b}{m_a^2 m_b} \hspace*{-3pt} \left( \hspace*{-4.4pt}\left( \hspace*{-3.4pt} -\frac{2 m_a m_b \hspace*{-0.4pt} (\hspace*{-0.4pt}3 m_a \hspace*{-3pt} + \hspace*{-2.4pt} 2 m_b \hspace*{-0.4pt})}{s - s_0} \hspace*{-1.8pt} - \hspace*{-1.8pt} \frac{15}{2} m_a \hspace*{-3pt} - \hspace*{-2pt} 5 m_b \hspace*{-3.2pt}\right) \hspace*{-2.9pt} S \hspace*{-2.2pt} + \hspace*{-2pt} \frac{32}{3} L \hspace*{-2.5pt}\right) \nonumber \\
&& \hspace*{32pt} +  i \frac{{\mathcal U}_a {\mathcal E}_b}{m_a m_b^2} \hspace*{-3pt} \left( \hspace*{-4.4pt}\left( \hspace*{-3.4pt} -\frac{2 m_a m_b \hspace*{-0.4pt} (\hspace*{-0.4pt}2 m_a \hspace*{-3pt} + \hspace*{-2.4pt} 3 m_b \hspace*{-0.4pt})}{s - s_0} \hspace*{-1.8pt} - \hspace*{-2.2pt} 5 m_a \hspace*{-3pt} - \hspace*{-1.9pt} \frac{15}{2} m_b \hspace*{-3.2pt}\right) \hspace*{-2.9pt} S \hspace*{-2.2pt} + \hspace*{-2pt} \frac{32}{3} L \hspace*{-2.5pt}\right) \nonumber \\
 && \hspace*{32pt} + \frac{S_a \cdot q S_b \cdot q - q^2 S_a \cdot S_b}{m_a m_b} \hspace*{-1pt} \left( \hspace*{-1pt} - \frac{2 m_a m_b}{s-s_0} - \frac{19}{2}\right) \hspace*{-1pt} (m_a + m_b) \hspace*{1pt} S  \nonumber \\
 && \hspace*{32pt} + \frac{S_a \cdot q S_b \cdot q}{m_a m_b} \ \frac{4}{3} \hspace*{1pt} L \hspace*{2pt} \Bigg] \nonumber \\
& + & \hspace*{-8pt} G \alpha \Bigg[{\mathcal U}_a {\mathcal U}_b \left( - (Z_a^2 m_b + Z_b^2 m_a) S - \frac{8(Z_a^2 m_b^2 + Z_b^2 m_a^2)}{3 m_a m_b} \hspace*{1pt} L \right) \nonumber \\
 && \hspace*{5pt} + i \frac{{\mathcal E}_a {\mathcal U}_b}{m_a^2 m_b} \bigg(\hspace*{-2pt} - \left(Z_a^2 m_b + \frac{3}{2} Z_b^2 m_a \hspace*{-1pt}\right) \hspace*{-1pt}S - \frac{4 (Z_a^2 m_b^2 + 3 Z_b^2 m_a^2)}{3 m_a m_b} \hspace*{1pt} L \hspace*{-1pt}\bigg) \nonumber \\
 && \hspace*{5pt} + i \frac{{\mathcal U}_a {\mathcal E}_b}{m_a m_b^2} \bigg(\hspace*{-2pt} - \left(\frac{3}{2} Z_a^2 m_b + Z_b^2 m_a \hspace*{-1pt}\right) \hspace*{-1pt}S - \frac{4 (3 Z_a^2 m_b^2 + Z_b^2 m_a^2)}{3 m_a m_b} \hspace*{1pt} L \hspace*{-1pt}\bigg) \nonumber \\
 && \hspace*{5pt} + \frac{S_a \hspace*{-3pt} \cdot \hspace*{-2.5pt} q S_b \hspace*{-3pt} \cdot \hspace*{-2.5pt} q \hspace*{-2pt} - \hspace*{-2pt} q^2 \hspace*{-0.5pt} S_a \hspace*{-3pt} \cdot \hspace*{-2.5pt} S_b}{m_a m_b} \hspace*{-2.7pt} \left(\hspace*{-4pt} -\frac{1}{2} \hspace*{-0.4pt} ( \hspace*{-1.2pt} Z_a^2 m_b \hspace*{-2.7pt} + \hspace*{-3pt} Z_b^2 m_a \hspace*{-0.7pt} ) \hspace*{-0.4pt} S \hspace*{-2.5pt} - \hspace*{-2pt} \frac{2 \hspace*{-0.4pt} (\hspace*{-1pt} Z_a^2 m_b^2 \hspace*{-2.7pt} + \hspace*{-3pt} Z_b^2 m_a^2\hspace*{-0.7pt} )}{3 m_a m_b} \hspace*{-0.7pt} L \hspace*{-3.5pt}  \right) \hspace*{-5.5pt} \Bigg] \nonumber \\
&+& \hspace*{-8pt} i 8\pi G  m_a m_b \, \alpha Z_a Z_b \hspace*{1.5pt} {L\over q^2} \sqrt{m_a m_b \over s-s_0} \left({\mathcal U}_a {\mathcal U}_b + \frac{3}{2} \, i \frac{{\mathcal U}_a {\mathcal E}_b}{m_a m_b^2} + \frac{3}{2} \, i \frac{{\mathcal E}_a {\mathcal U}_b}{m_a^2 m_b}\right. \nonumber \\
 && \hspace*{147pt} \left. + \frac{S_a \cdot q S_b \cdot q - \frac{1}{2} q^2 S_a \cdot S_b}{m_a m_b} \right).
\quad \label{eq_mix_Mhh_rel}
\end{eqnarray}
Finally, we take the nonrelativistic limit as before, yielding
\begin{eqnarray}
{}^{\frac{1}{2} \frac{1}{2}}{\cal M}_{tot}^{(2)}(\vec q)\hspace*{-7pt} &\simeq\hspace*{-7pt}&\Bigg[ G \alpha Z_a Z_b
\Big(\hspace*{-3pt} - \hspace*{-1pt} 6 (m_a \hspace*{-1pt} + \hspace*{-1pt} m_b)S \hspace*{-1pt} + \hspace*{-1pt} 12 L \Big) \hspace*{-1pt} + i 8\pi G  m_a m_b \, \alpha Z_a Z_b \hspace*{1.5pt} {L\over q^2} \frac{m_r}{p_0} \nonumber \\
&& + \hspace*{1pt} G \alpha \hspace*{-1pt} \left( \hspace*{-2pt} - (Z_a^2 m_b \hspace*{-0.4pt} + \hspace*{-0.4pt} Z_b^2 m_a) S - \frac{8(Z_a^2 m_b^2 \hspace*{-0.4pt} + \hspace*{-0.4pt} Z_b^2 m_a^2)}{3 m_a m_b} \hspace*{1pt} L \right) \hspace*{-3.5pt} \Bigg]
\chi_f^{a\dagger}\chi_i^a \, \chi_f^{b\dagger}\chi_i^b \nonumber\\
&+\hspace*{-7pt}&\Bigg[\hspace*{-0.7pt} \frac{G \alpha Z_a Z_b}{m_a m_b} \hspace*{-3pt} \left(\hspace*{-3.5pt} - \frac{9m_a^3 \hspace*{-2.5pt} + \hspace*{-2.5pt} 18 m_a^2 m_b \hspace*{-2.5pt} + \hspace*{-2.5pt} 13 m_a m_b^2
 \hspace*{-2.5pt} + \hspace*{-2.5pt} 3 m_b^3}{m_a \hspace*{-1pt} + \hspace*{-1pt} m_b} S \hspace*{-1.5pt}
+ \hspace*{-1.5pt} \frac{32 m_a \hspace*{-2.5pt} + \hspace*{-2.5pt} 14 m_b}{3}  L \hspace*{-3pt} \right)\nonumber\\
&&- \frac{2 G m_a m_b \hspace*{1pt} \alpha Z_a Z_b (3 m_a + 2 m_b)}{(m_a + m_b)} \left(- i \frac{2 \pi
L}{p_0 q^2} + \frac{S}{p_0^2} \right) \nonumber \\
&& + \frac{G \alpha}{m_a m_b} \bigg( \hspace*{-3.5pt} - \hspace*{-1pt} \frac{1}{2}\Big(Z_a^2 (2 m_a m_b + m_b^2) + Z_b^2 (3 m_a^2 + 2 m_a m_b)\Big)S \nonumber \\
&& \hspace*{44pt} - \frac{4(Z_a^2 m_b^2 \hspace*{-1.5pt} + \hspace*{-1.5pt} Z_b^2 (3 m_a^2 \hspace*{-1.5pt} + \hspace*{-1.5pt} 2 m_a m_b) \hspace*{-0.5pt} )}{3 m_b} L \hspace*{-1.7pt} \bigg) \hspace*{-1.9pt} \Bigg]
{i\over m_a}\vec{S}_a \hspace*{-1.5pt} \cdot \hspace*{-1pt} \vec{p} \hspace*{-0.5pt} \times \hspace*{-0.5pt} \vec{q} \ \chi_f^{b\dagger} \chi_i^b  \nonumber \\
&+\hspace*{-7pt}&\Bigg[\hspace*{-0.7pt} \frac{G \alpha Z_a Z_b}{m_a m_b} \hspace*{-3pt} \left(\hspace*{-3.5pt} - \frac{3m_a^3 \hspace*{-2.5pt} + \hspace*{-2.5pt} 13 m_a^2 m_b \hspace*{-2.5pt} + \hspace*{-2.5pt} 18 m_a m_b^2
 \hspace*{-2.5pt} + \hspace*{-2.5pt} 9 m_b^3}{m_a \hspace*{-1pt} + \hspace*{-1pt} m_b} S \hspace*{-1.5pt}
+ \hspace*{-1.5pt} \frac{14 m_a \hspace*{-2.5pt} + \hspace*{-2.5pt} 32 m_b}{3}  L \hspace*{-3pt} \right)\nonumber\\
&&- \frac{2 G m_a m_b \hspace*{1pt} \alpha Z_a Z_b (2 m_a + 3 m_b)}{(m_a + m_b)} \left(- i \frac{2 \pi
L}{p_0 q^2} + \frac{S}{p_0^2} \right) \nonumber \\
&& + \frac{G \alpha}{m_a m_b} \bigg( \hspace*{-3.5pt} - \hspace*{-1pt} \frac{1}{2}\Big(Z_a^2 (2 m_a m_b + 3 m_b^2) + Z_b^2 (m_a^2 + 2 m_a m_b)\Big)S \nonumber \\
&& \hspace*{44pt} - \frac{4(Z_a^2(2 m_a m_b \hspace*{-1.5pt} + \hspace*{-1.5pt} 3 m_b^2) \hspace*{-1.5pt} + \hspace*{-1.5pt} Z_b^2 m_a^2)}{3 m_a} L \hspace*{-1.7pt} \bigg) \hspace*{-1.9pt}\Bigg]
 \chi_f^{a\dagger}\chi_i^a \, {i\over m_b}\vec{S}_b \hspace*{-1.5pt} \cdot \hspace*{-1pt} \vec{p} \hspace*{-0.5pt} \times \hspace*{-0.5pt} \vec{q} \nonumber \\
&-\hspace*{-7pt}& G \alpha Z_a Z_b \ \frac{2(5 m_a^2 + 9 m_a m_b + 5 m_b^2)}{m_a + m_b} \hspace*{1pt} S \ \frac{\vec S_a \cdot \vec q \, \vec S_b \cdot \vec q - \vec q^{\hspace*{1.4pt}2} \vec S_a \cdot \vec S_b}{m_a m_b} \nonumber \\
&+\hspace*{-7pt}& G \alpha Z_a Z_b \ \frac{4}{3} \hspace*{1pt} L \ \frac{\vec S_a \cdot \vec q \, \vec S_b \cdot \vec q}{m_a m_b} \nonumber \\
& - \hspace*{-7pt}& \frac{2 G m_a^2 m_b^2 \, \alpha Z_a Z_b}{m_a + m_b} \, \frac{S}{p_0^2} \,  \frac{\vec S_a \cdot \vec q \, \vec S_b \cdot \vec q - \vec q^{\hspace*{1.4pt}2} \vec S_a \cdot \vec S_b}{m_a m_b} \nonumber \\
& - \hspace*{-7pt}& \frac{2 G m_a^2 m_b^2 \, \alpha Z_a Z_b}{m_a + m_b} \left(-i \frac{4 \pi L}{p_0 q^2}\right)
\frac{\vec S_a \cdot \vec q \, \vec S_b \cdot \vec q - \frac{1}{2}\vec q^{\hspace*{1.4pt}2} \vec S_a \cdot \vec S_b}{m_a m_b} \nonumber \\
& + \hspace*{-7pt}& G \alpha \Bigg( \hspace*{-3pt} - \frac{1}{2} (Z_a^2 m_b \hspace*{-1.5pt} + \hspace*{-1.5pt} Z_b^2 m_a) S \nonumber \\
&& \hspace*{25pt} - \hspace*{-1pt} \frac{2(Z_a^2 m_b^2 \hspace*{-1.5pt} + \hspace*{-1.5pt} Z_b^2 m_a^2)}{3 m_a m_b} \hspace*{1pt} L \Bigg)  \frac{\vec S_a \hspace*{-1.5pt} \cdot \hspace*{-1pt} \vec q \, \vec S_b \hspace*{-1.5pt} \cdot \hspace*{-1pt} \vec q - \vec q^{\hspace*{1.4pt}2} \vec S_a \hspace*{-1.5pt} \cdot \hspace*{-1.5pt} \vec S_b}{m_a m_b} \hspace*{-1pt} .\nonumber \\
 \quad \label{eq_mix_Mhh_nr}
\end{eqnarray}
The amplitude for spin-1/2 -- spin-1/2 scattering now involves a spin
independent component, two spin-orbit coupling pieces (one for each
spin) which are symmetric under $a \leftrightarrow b$ and new
spin-spin coupling terms. Comparing with the corresponding
expression for the spin-0 -- spin-1/2 scattering amplitude---Eq.
(\ref{eq_mix_Mh_nr})---we observe that the kinematic forms
multiplying {\it both} the spin-independent and spin-orbit terms are
identical. This is the universality which we expected and confirms
the correctness of the individual diagram calculations.

Of course, before we can generate a proper potential we must, as
before, subtract off the imaginary $i L /(q^2 p_0)$ and the real $S
/ p_0^2$ pieces by iterating the lowest order electromagnetic and
gravitational potentials.  In the electromagnetic case the leading
order potential for spin-1/2 -- spin-1/2 scattering reads \cite{hrem}
\begin{equation}
\left< \hspace*{-1pt} \vec p_f \hspace*{-1pt} \left| {}^{{1\over 2}{1\over 2}} \hat V^{(1)}_C
\right| \hspace*{-1pt} \vec p_i \right> =
 \left< \hspace*{-1pt} \vec p_f \hspace*{-1pt} \left| {}^{{1\over 2}{1\over 2}} \hat V^{(1)}_{C, \hspace*{1pt} S-I} \right| \hspace*{-1pt} \vec p_i \right>
 + \left< \hspace*{-1pt} \vec p_f \hspace*{-1pt} \left| {}^{{1\over 2}{1\over 2}} \hat V^{(1)}_{C, \hspace*{1pt} S-O} \right| \hspace*{-1pt} \vec p_i \right>
 + \left< \hspace*{-1pt} \vec p_f \hspace*{-1pt} \left| {}^{{1\over 2}{1\over 2}} \hat V^{(1)}_{C, \hspace*{1pt} S-S} \right| \hspace*{-1pt} \vec p_i \right>
\end{equation}
with the components
\begin{eqnarray}
 \left<\vec p_f \left| {}^{{1\over 2}{1\over 2}} \hat V^{(1)}_{C, \hspace*{1pt} S-I} \right| \vec p_i \, \right>
 &=&{c_C^2\over \vec{q}^{\hspace*{1.4pt}2}} \, \chi_f^{a\dagger}\chi_i^a \, \chi_f^{b\dagger}\chi_i^b \nonumber\\
 \left<\vec p_f \left| {}^{{1\over 2}{1\over 2}} \hat V^{(1)}_{C, \hspace*{1pt} S-O} \right| \vec p_i \, \right>
 &=&{c_C^2\over \vec{q}^{\hspace*{1.4pt}2}}{2 m_a+ m_b\over 2m_am_b}\, \frac{i}{m_a}\vec{S}_a\cdot\vec{p}\times\vec{q} \, \chi_f^{b\dagger}\chi_i^b\nonumber\\
 &+&{c_C^2\over \vec{q}^{\hspace*{1.4pt}2}}{m_a+2m_b\over 2m_am_b} \, \chi_f^{a\dagger}\chi_i^a \, \frac{i}{m_b}\vec{S}_b\cdot\vec{p}\times\vec{q} \nonumber\\
 \left<\vec p_f \left| {}^{{1\over 2}{1\over 2}} \hat V^{(1)}_{C, \hspace*{1pt} S-S} \right| \vec p_i \, \right>
 &=&{c_C^2\over \vec{q}^{\hspace*{1.4pt}2}} {1\over m_am_b} \, \vec{S}_a\cdot\vec{q} \, \vec{S}_b\cdot\vec{q}
\end{eqnarray}
and where $q = p_i - p_f$ and $p = \frac{1}{2} (p_i + p_f)$ when identifying $\vec p_i$ with $\vec p_1$ and $\vec p_f$ with $\vec p_2$.
The leading order gravitational potential obtained in \cite{hrgr} is
\begin{equation}
\left< \hspace*{-1pt} \vec p_f \hspace*{-1pt} \left| {}^{{1\over 2}{1\over 2}} \hat V^{(1)}_G
\right| \hspace*{-1pt} \vec p_i \right> =
 \left< \hspace*{-1pt} \vec p_f \hspace*{-1pt} \left| {}^{{1\over 2}{1\over 2}} \hat V^{(1)}_{G, \hspace*{1pt} S-I} \right| \hspace*{-1pt} \vec p_i \right>
 + \left< \hspace*{-1pt} \vec p_f \hspace*{-1pt} \left| {}^{{1\over 2}{1\over 2}} \hat V^{(1)}_{G, \hspace*{1pt} S-O} \right| \hspace*{-1pt} \vec p_i \right>
 + \left< \hspace*{-1pt} \vec p_f \hspace*{-1pt} \left| {}^{{1\over 2}{1\over 2}} \hat V^{(1)}_{G, \hspace*{1pt} S-S} \right| \hspace*{-1pt} \vec p_i \right>
\end{equation}
with the components
\begin{eqnarray}
 \left<\vec p_f \left| {}^{{1\over 2}{1\over 2}} \hat V^{(1)}_{G, \hspace*{1pt} S-I} \right| \vec p_i \, \right>
 &=&{c_G^2\over \vec{q}^{\hspace*{1.4pt}2}} \, \chi_f^{a\dagger}\chi_i^a \, \chi_f^{b\dagger}\chi_i^b \nonumber\\
 \left<\vec p_f \left| {}^{{1\over 2}{1\over 2}} \hat V^{(1)}_{G, \hspace*{1pt} S-O} \right| \vec p_i \, \right>
 &=&{c_G^2\over \vec{q}^{\hspace*{1.4pt}2}}{4 m_a + 3 m_b\over 2m_am_b}\, \frac{i}{m_a}\vec{S}_a\cdot\vec{p}\times\vec{q} \, \chi_f^{b\dagger}\chi_i^b\nonumber\\
 &+&{c_G^2\over \vec{q}^{\hspace*{1.4pt}2}}{3 m_a + 4 m_b\over 2m_am_b} \, \chi_f^{a\dagger}\chi_i^a \, \frac{i}{m_b}\vec{S}_b\cdot\vec{p}\times\vec{q} \nonumber\\
 \left<\vec p_f \left| {}^{{1\over 2}{1\over 2}} \hat V^{(1)}_{G, \hspace*{1pt} S-S} \right| \vec p_i \, \right>
 &=&{c_G^2\over \vec{q}^{\hspace*{1.4pt}2}} {1\over m_am_b} \, \vec{S}_a\cdot\vec{q} \, \vec{S}_b\cdot\vec{q}.
\end{eqnarray}

Note that in both the electromagnetic and gravitational cases the
spin-independent and spin-orbit pieces of the lowest order potential
are identical to the results found in the case of spin-0 -- spin-1/2
scattering, as required by universality.  This means that when the
second order Born amplitude is evaluated the calculation of these
pieces will go through as before and that the offending $i L / (q^2
p_0)$ and $S/p_0^2$ pieces will be removed just as in the
spin-0 -- spin-1/2 case.  However, both lowest order potentials also
contain a new spin-spin component and it is this piece which will
generate a spin-spin term in the second Born amplitude when iterated
together with the leading spin-independent piece of
$V_{C,G}^{(1)}(\vec{r})$. (Note that a spin-spin component will also
arise from the product of spin-orbit components of the lowest order
gravitational and electromagnetic potentials, but this piece is
higher order in $p_0^2$ and is not relevant for our calculation.)
The leading spin-spin second Born term is then
\begin{eqnarray}
{}^{{1\over 2}{1\over 2}} {\rm Amp}^{(2)}_{S-S}(\vec q)
\hspace*{-7pt}{} &=\hspace*{-7pt}&- \int{d^3\ell\over
(2\pi)^3} \, \frac{\left<\vec p_f \left| {}^{{1\over 2}{1\over 2}} \hat V^{(1)}_{G, \hspace*{1pt} S-I} \right| \vec \ell \, \right> \left<\vec \ell \left| {}^{{1\over 2}{1\over 2}} \hat V^{(1)}_{C, \hspace*{1pt} S-S} \right| \vec p_i \right>}{\frac{p_0^2}{2 m_r} - \frac{\ell^2}{2 m_r} + i \epsilon} \nonumber\\
&&- \int{d^3\ell\over (2\pi)^3} \, \frac{\left<\vec p_f \left| {}^{{1\over 2}{1\over 2}} \hat V^{(1)}_{C, \hspace*{1pt} S-S} \right| \vec \ell \, \right> \left<\vec \ell \left| {}^{{1\over 2}{1\over 2}} \hat V^{(1)}_{G, \hspace*{1pt} S-I} \right| \vec p_i \right>}{\frac{p_0^2}{2 m_r} - \frac{\ell^2}{2 m_r} + i \epsilon} \nonumber\\
&&- \int{d^3\ell\over (2\pi)^3} \, \frac{\left<\vec p_f \left| {}^{{1\over 2}{1\over 2}} \hat V^{(1)}_{G, \hspace*{1pt} S-S} \right| \vec \ell \, \right> \left<\vec \ell \left| {}^{{1\over 2}{1\over 2}} \hat V^{(1)}_{C, \hspace*{1pt} S-I} \right| \vec p_i \right>}{\frac{p_0^2}{2 m_r} - \frac{\ell^2}{2 m_r} + i \epsilon} \nonumber\\
&&- \int{d^3\ell\over (2\pi)^3} \, \frac{\left<\vec p_f \left| {}^{{1\over 2}{1\over 2}} \hat V^{(1)}_{C, \hspace*{1pt} S-I} \right| \vec \ell \, \right> \left<\vec \ell \left| {}^{{1\over 2}{1\over 2}} \hat V^{(1)}_{G, \hspace*{1pt} S-S} \right| \vec p_i \right>}{\frac{p_0^2}{2 m_r} - \frac{\ell^2}{2 m_r} + i \epsilon} \nonumber\\
&=\hspace*{-7pt}&{1\over m_am_b} \, S_a^r \, S_b^s \nonumber\\
&& \left(\hspace*{-2pt}i \hspace*{-3.2pt} \int \hspace*{-3.2pt}
{d^3\ell\over (2\pi)^3} {c_G^2\over | \vec{\ell} \hspace*{-1.1pt} -
\hspace*{-1.1pt} \vec{p}_f \hspace*{1pt}|^2 \hspace*{-1.1pt} +
\hspace*{-1.2pt} \lambda^2}{i \over {p_0^2\over 2m_r}
\hspace*{-1.1pt} - \hspace*{-1.1pt} {\ell^2\over 2m_r}
\hspace*{-1.1pt} + \hspace*{-1.1pt} i\epsilon}{c_C^2 (p_i
\hspace*{-1.1pt} - \hspace*{-1.1pt} \ell)^r (p_i \hspace*{-1.1pt} -
\hspace*{-1.1pt} \ell)^s\over
|\vec{p}_i \hspace*{-1.1pt} - \hspace*{-1.1pt} \vec{\ell}|^2 \hspace*{-1.1pt} + \hspace*{-1.2pt} \lambda^2}\right.\nonumber\\
&&\left. \hspace*{-3.6pt}+ i \hspace*{-3.2pt} \int \hspace*{-3.2pt}
{d^3\ell\over (2\pi)^3} {c_C^2 (\ell \hspace*{-1.1pt} -
\hspace*{-1.2pt} p_f)^r (\ell \hspace*{-1.1pt} - \hspace*{-1.2pt}
p_f)^s \over |\vec{\ell} \hspace*{-1.1pt} - \hspace*{-1.1pt}
\vec{p}_f \hspace*{1pt}|^2 \hspace*{-1.1pt} + \hspace*{-1.2pt}
\lambda^2}{i \over {p_0^2\over 2m_r} \hspace*{-1.1pt} -
\hspace*{-1.1pt} {\ell^2\over 2m_r} \hspace*{-1.1pt} +
\hspace*{-1.1pt} i\epsilon}{c_G^2\over
|\vec{p}_i \hspace*{-1.1pt} - \hspace*{-1.1pt} \vec{\ell}|^2 \hspace*{-1.1pt} + \hspace*{-1.2pt}\lambda^2}\hspace*{-3pt}\right)\nonumber\\
&&\left. \hspace*{-3.6pt}+ i \hspace*{-3.2pt} \int \hspace*{-3.2pt}
{d^3\ell\over (2\pi)^3} {c_G^2 (\ell \hspace*{-1.1pt} -
\hspace*{-1.2pt} p_f)^r (\ell \hspace*{-1.1pt} - \hspace*{-1.2pt}
p_f)^s \over |\vec{\ell} \hspace*{-1.1pt} - \hspace*{-1.1pt}
\vec{p}_f \hspace*{1pt}|^2 \hspace*{-1.1pt} + \hspace*{-1.2pt}
\lambda^2}{i \over {p_0^2\over 2m_r} \hspace*{-1.1pt} -
\hspace*{-1.1pt} {\ell^2\over 2m_r} \hspace*{-1.1pt} +
\hspace*{-1.1pt} i\epsilon}{c_C^2\over
|\vec{p}_i \hspace*{-1.1pt} - \hspace*{-1.1pt} \vec{\ell}|^2 \hspace*{-1.1pt} + \hspace*{-1.2pt}\lambda^2}\hspace*{-3pt}\right)\nonumber\\
&&\left. \hspace*{-3.6pt}+ i \hspace*{-3.2pt} \int \hspace*{-3.2pt}
{d^3\ell\over (2\pi)^3} {c_C^2\over | \vec{\ell} \hspace*{-1.1pt} -
\hspace*{-1.1pt} \vec{p}_f \hspace*{1pt}|^2 \hspace*{-1.1pt} +
\hspace*{-1.2pt} \lambda^2}{i \over {p_0^2\over 2m_r}
\hspace*{-1.1pt} - \hspace*{-1.1pt} {\ell^2\over 2m_r}
\hspace*{-1.1pt} + \hspace*{-1.1pt} i\epsilon}{c_G^2 (p_i
\hspace*{-1.1pt} - \hspace*{-1.1pt} \ell)^r (p_i \hspace*{-1.1pt} -
\hspace*{-1.1pt} \ell)^s\over
|\vec{p}_i \hspace*{-1.1pt} - \hspace*{-1.1pt} \vec{\ell}|^2 \hspace*{-1.1pt} + \hspace*{-1.2pt} \lambda^2}\right.\nonumber\\
&\stackrel{\lambda\rightarrow 0}{\longrightarrow} \hspace*{-8pt}&{2\over m_a m_b} \Bigg[\left(\vec{S}_a\cdot\vec{p_i} \, \vec{S}_b\cdot\vec{p_i} + \vec{S}_a\cdot\vec{p_f} \, \vec{S}_b\cdot\vec{p_f} \right) H\nonumber\\
&& \hspace*{32pt} - \vec S_a \cdot (\vec p_i + p_f) \vec S_b \cdot \vec H - \vec S_a \cdot \vec H \, \vec S_b \cdot (\vec p_i + p_f) \nonumber\\
&& \hspace*{32pt} + 2 \, S_a^r S_b^s \, H^{rs}\Bigg]\nonumber\\
&=\hspace*{-7pt}& - \frac{2 G m_a^2 m_b^2 \hspace*{1pt} \alpha Z_a Z_b}{m_a + m_b} \, \frac{S}{p_0^2} \,  \frac{\vec S_a \cdot \vec q \, \vec S_b \cdot \vec q - \vec q^{\hspace*{1.4pt}2} \vec S_a \cdot \vec S_b}{m_a m_b}\nonumber\\
&& - \frac{2 G m_a^2 m_b^2 \hspace*{1pt} \alpha Z_a Z_b}{m_a + m_b} \left(-i \frac{4 \pi L}{p_0
q^2}\right) \frac{\vec S_a \cdot \vec q \, \vec S_b \cdot \vec q -
\frac{1}{2}\vec q^{\hspace*{1.4pt}2} \vec S_a \cdot \vec S_b}{m_a m_b}. \nonumber \\ \quad \label{eq_iteration_spinspin_em}
\end{eqnarray}

Comparing with the second order scattering amplitude Eq. (\ref{eq_mix_Mhh_nr})
we see that the iterated result Eq. (\ref{eq_iteration_spinspin_em}) is of  precisely the
right form to eliminate the $iL/(q^2p_0)$ and $S/p_0^2$ pieces
of the spin-spin correlation.  The resulting second order potential
is then well-defined and is of the form
\begin{eqnarray}
{}^{\frac{1}{2}\frac{1}{2}}V_{CG}^{(2)}(\vec r) \hspace*{-6pt} &=&-\int{d^3q\over
(2\pi)^3}e^{-i\vec{q}\cdot\vec{r}}\left[{}^{{1\over 2}{1\over 2}}{\cal M}^{(2)}_{tot}(\vec{q})- \, {}^{{1\over 2}{1\over 2}}{\rm Amp}_{CG}^{(2)}(\vec{q})\right]\nonumber\\
& \simeq \hspace*{-6pt} & G \alpha \Bigg[\frac{1}{2} \frac{ (Z_a^2 m_b + Z_b^2 m_a)}{r^2} + 3 \frac{Z_a Z_b (m_a + m_b)}{r^2} \nonumber \\
&& {} \hspace*{15pt} - \frac{4 \hbar}{{3 \pi r^3}} \left(Z_b^2 \frac{m_a}{m_b} + Z_a^2 \frac{m_b}{m_a}\right) + \frac{6 Z_a Z_b \hbar}{\pi r^3} \Bigg] \chi_f^{a\dagger}\chi_i^a \, \chi_f^{b\dagger}\chi_i^b \nonumber \\
&+\hspace*{-6pt} & \Bigg[- \frac{Z_a Z_b G \alpha(9 m_a^3 + 18 m_a^2 m_b + 13 m_a m_b^2 + 3 m_b^3)}{m_a^2 m_b (m_a + m_b) \hspace*{1pt} r^4}  \nonumber \\
&& {} \hspace*{-1pt} - \frac{G \alpha(Z_a^2(2 m_a m_b + m_b^2) + Z_b^2 (3 m_a^2 + 2 m_a m_b))}{2 m_a^2 m_b \hspace*{1pt} r^4} \nonumber \\
&& {} \hspace*{-1pt} - \frac{Z_a Z_b G \alpha \hbar (16 m_a + 7 m_b)}{\pi m_a^2 m_b \hspace*{1pt} r^5} \nonumber \\
&& {} \hspace*{-1pt} + \frac{2 G \alpha \hbar (Z_a^2 m_b^2 + Z_b^2 (3 m_a^2 + 2 m_a m_b))}{\pi m_a^2 m_b^2 \hspace*{1pt} r^5} \hspace*{1pt} \Bigg] \hspace*{2pt} \vec L \cdot \vec S_a \hspace*{3pt} \chi_f^{b\dagger} \hspace*{-0.5pt} \chi_i^b  \nonumber \\
&+\hspace*{-6pt} & \Bigg[- \frac{Z_a Z_b G \alpha(3 m_a^3 + 13 m_a^2 m_b + 18 m_a m_b^2 + 9 m_b^3)}{m_a m_b^2 (m_a + m_b) \hspace*{1pt} r^4}  \nonumber \\
&& {} \hspace*{-1pt} - \frac{G \alpha(Z_a^2(2 m_a m_b + 3 m_b^2) + Z_b^2 (m_a^2 + 2 m_a m_b))}{2 m_a m_b^2 \hspace*{1pt} r^4} \nonumber \\
&& {} \hspace*{-1pt} - \frac{Z_a Z_b G \alpha \hbar (7 m_a + 16 m_b)}{\pi m_a m_b^2 \hspace*{1pt} r^5} \nonumber \\
&& {} \hspace*{-1pt} + \frac{2 G \alpha \hbar (Z_a^2(2 m_a m_b + 3 m_b^2) + Z_b^2 m_a^2)}{\pi m_a^2 m_b^2 \hspace*{1pt} r^5} \hspace*{1pt}\Bigg] \hspace*{2pt} \chi_f^{a\dagger} \hspace*{-0.5pt} \chi_i^a \hspace*{2pt} \vec L \cdot \vec S_b.  \nonumber \\
&& {} \hspace*{-1pt} - \frac{8 Z_a Z_b G \alpha(5 m_a^2 \hspace*{-1pt} + \hspace*{-1pt} 9 m_a m_b \hspace*{-1pt} + \hspace*{-1pt} 5 m_b^2)}{m_a m_b (m_a + m_b) \hspace*{1pt} r^4} \hspace*{-1pt} \left(\hspace*{-1pt} \vec S_a \hspace*{-1pt} \cdot \hspace*{-0.5pt} \vec r \, \vec S_b \hspace*{-1pt} \cdot \hspace*{-0.5pt} \vec r \, / r^2 - \frac{1}{2} \hspace*{1pt} \vec S_a \hspace*{-1pt} \cdot \hspace*{-1pt} \vec S_b \hspace*{-1pt} \right) \nonumber \\
&& {} \hspace*{-1pt} - \frac{2 G \alpha(Z_a^2 m_b + Z_b^2 m_a)}{m_a m_b \hspace*{1pt} r^4} \left(\vec S_a \cdot \vec r \, \vec S_b \cdot \vec r \, / r^2 - \frac{1}{2} \hspace*{1pt} \vec S_a \cdot \vec S_b\right) \nonumber \\
&& {} \hspace*{-1pt} - \frac{10 Z_a Z_b G \alpha \hbar}{\pi m_a m_b \hspace*{1pt} r^5} \left(\vec S_a \cdot \vec r \, \vec S_b \cdot \vec r \, / r^2 - \frac{1}{5} \hspace*{1pt} \vec S_a \cdot \vec S_b\right) \nonumber \\
&& {} \hspace*{-1pt} + \frac{5 G \alpha \hbar \hspace*{0.5pt} (Z_a^2 m_b^2 + Z_b^2 m_a^2)}{\pi m_a^2 m_b^2 \hspace*{1pt} r^5} \left(\vec S_a \cdot \vec r \, \vec S_b \cdot \vec r \, / r^2 - \frac{3}{5} \hspace*{1pt} \vec S_a \cdot \vec S_b\right)  \quad \label{eq_nlo_mixhh}
\end{eqnarray}
Comparing with the earlier result found in Eq. (\ref{eq:hd}) for
spin-0 -- spin-1/2 scattering, we confirm universality for the
spin-independent and spin-orbit components together with a new
spin-spin piece which itself is presumably universal.

\section{Spin-0 -- Spin-1 Scattering}

Before closing we note that we have also evaluated the case of
scattering of a spinless particle $a$ from a particle $b$ of unit
spin as a further check of universality and of the complications
which arise from the existence of quadrupole effects.  Using the
vertices given in \cite{hrem}, \cite{hrgr} and in Appendix A---note that the electromagnetic
vertices for the unit spin system use the "natural" value
$g=2$ \cite{Holstein:2006wi}---we find the results
\begin{eqnarray}
{}^1{\cal M}^{(2)}_{\ref{fig_diags}a}(q) \hspace*{-5pt} &=\hspace*{-5pt} & G \alpha Z_a Z_b \Bigg[-12 L \hspace*{1pt} \epsilon_f^{*b} \cdot \epsilon_i^b
- \frac{9 L}{m_a m_b} (\epsilon_f^{b*}\cdot q \epsilon_i^b \cdot p_1 - \epsilon_f^{b*} \cdot p_1 \epsilon_i^b\cdot q) \Bigg] \nonumber\\
{}^1{\cal M}^{(2)}_{\ref{fig_diags}b}(q) \hspace*{-5pt} &=\hspace*{-5pt} & G \alpha Z_a Z_b \Bigg[(10 L + 4 m_a S) \epsilon_f^{*b} \cdot \epsilon_i^b \nonumber\\
&&\hspace*{41pt} + \frac{8 L + 3 m_a S}{m_a m_b} (\epsilon_f^{b*}\cdot q \epsilon_i^b \cdot p_1 - \epsilon_f^{b*} \cdot p_1 \epsilon_i^b\cdot q) \nonumber\\
&&\hspace*{41pt} + \left(\frac{m_b^2}{2 m_a^2} L + \frac{m_b^2}{4 m_a} S \right) \frac{1}{m_b^2} \epsilon_f^{b*}\cdot q \epsilon_i^b \cdot q \Bigg] \nonumber\\
{}^1{\cal M}^{(2)}_{\ref{fig_diags}c}(q) \hspace*{-5pt} &=\hspace*{-5pt} & G \alpha Z_a Z_b \Bigg[(12 L + 8 m_b S) \epsilon_f^{*b} \cdot \epsilon_i^b \nonumber\\
&&\hspace*{41pt} + \frac{5 L + 2 m_b S}{m_a m_b} (\epsilon_f^{b*}\cdot q \epsilon_i^b \cdot p_1 - \epsilon_f^{b*} \cdot p_1 \epsilon_i^b\cdot q) \nonumber\\
&&\hspace*{41pt} + \left(3 L + 3 m_b S \right) \frac{1}{m_b^2} \epsilon_f^{b*}\cdot q \epsilon_i^b \cdot q \Bigg] \nonumber\\
{}^1{\cal M}^{(2)}_{\ref{fig_diags}d}(q) \hspace*{-5pt} &=\hspace*{-5pt} & G \alpha Z_a Z_b \Bigg[
\frac{4 m_a m_b} {q^2} L \bigg( \hspace*{-1.5pt} 2 \epsilon_f^{*b} \hspace*{-1pt} \cdot \hspace*{-0.5pt} \epsilon_i^b  \hspace*{-0.5pt}
+ \frac{3}{m_a m_b} (\epsilon_f^{b*} \hspace*{-1pt} \cdot \hspace*{-0.5pt} q \epsilon_i^b \hspace*{-1pt} \cdot \hspace*{-0.5pt} p_1 \hspace*{-1pt} - \epsilon_f^{b*} \hspace*{-1pt} \cdot \hspace*{-0.5pt} p_1 \epsilon_i^b\cdot \hspace*{-0.5pt} q) \nonumber \\
&& \hspace*{100pt} - \frac{1}{m_b^2} \epsilon_f^{b*}\cdot q \epsilon_i^b \cdot q\bigg) \nonumber \\
&&\hspace*{41pt} + \frac{S}{s-s_0} \bigg( \hspace*{-2.5pt} (4 m_a + 6 m_b) (\epsilon_f^{b*}\cdot q \epsilon_i^b \cdot p_1 - \epsilon_f^{b*} \cdot p_1 \epsilon_i^b\cdot q)\nonumber\\
&&\hspace*{88pt} - \frac{m_a (3 m_a + 5 m_b)}{m_b} \epsilon_f^{b*}\cdot q \epsilon_i^b \cdot q \bigg)\nonumber\\
&&\hspace*{41pt} + \bigg(\frac{12 m_a^2 - 3 m_a m_b + 21 m_b^2}{3 m_a m_b} \hspace*{1pt} L + (7 m_a + 5 m_b) S \bigg) \epsilon_f^{*b} \cdot \epsilon_i^b \nonumber\\
&& \hspace*{41pt} - \frac{20}{3} \hspace*{1pt} L \hspace*{1pt} \,  \frac{1}{m_a m_b} \hspace*{1pt} \epsilon_f^{b*}\cdot p_1 \epsilon_i^b \cdot p_1 \nonumber \\
&& \hspace*{41pt} + \bigg(\frac{78 m_a^2 + 4 m_a m_b + 99 m_b^2}{12 m_a m_b} \hspace*{1pt} L + \left(\frac{17}{2} m_a + 11 m_b\right) \hspace*{-1pt} S \bigg) \nonumber \\
&& \hspace*{59pt} \times \frac{1}{m_a m_b} (\epsilon_f^{b*}\cdot q \epsilon_i^b \cdot p_1 - \epsilon_f^{b*} \cdot p_1 \epsilon_i^b\cdot q) \nonumber \\
&& \hspace*{41pt} + \frac{10}{3} \hspace*{1pt} L \, \frac{1}{m_a m_b} (\epsilon_f^{b*} \hspace*{-1.5pt} \cdot \hspace*{-0.8pt} q \epsilon_i^b \hspace*{-1.5pt} \cdot \hspace*{-0.8pt} p_1 \hspace*{-1pt} + \hspace*{-0.7pt} \epsilon_f^{b*} \hspace*{-1.5pt} \cdot \hspace*{-0.8pt} p_1 \epsilon_i^b \hspace*{-1.5pt} \cdot \hspace*{-0.8pt} q)\nonumber \\
&& \hspace*{41pt} - \bigg(\hspace*{-2.5pt}  \bigg( \hspace*{-1pt} \frac{7 m_a}{3 m_b} \hspace*{-1.5pt} + \hspace*{-1.5pt} \frac{1}{6} \hspace*{-1.5pt} + \hspace*{-1.5pt} \frac{16 m_b}{3 m_a} \hspace*{-1.5pt} + \hspace*{-1.5pt} \frac{m_b^2}{4m_a^2}\bigg) L \nonumber \\
&& \hspace*{55pt} + \hspace*{-1pt} \bigg(\hspace*{-1pt} 2 m_a \hspace*{-1.5pt} + \hspace*{-1.5pt} \frac{11}{2} m_b \hspace*{-1.5pt} + \hspace*{-1.5pt} \frac{m_b^2}{8 m_a} \hspace*{-1.5pt} \bigg) S \hspace*{-1pt}  \bigg)
  \times \frac{1}{m_b^2} \epsilon_f^{b*}\cdot q \epsilon_i^b \cdot q  \Bigg] \nonumber \\
&+& \hspace*{-5pt} i 8\pi G  m_a m_b \, \alpha Z_a Z_b \hspace*{1.5pt} {L\over q^2} \sqrt{m_a m_b \over s-s_0} \hspace*{-0.1pt} \Bigg( \hspace*{-4pt}
- \hspace*{-0.5pt} \epsilon_f^{*b} \hspace*{-1.5pt} \cdot \hspace*{-0.5pt} \epsilon_i^b\nonumber \\
&& \hspace*{146pt} - \frac{3}{2} \, \frac{1}{m_a m_b} (\epsilon_f^{b*} \hspace*{-2pt} \cdot \hspace*{-1.3pt} q \epsilon_i^b \hspace*{-2pt} \cdot \hspace*{-1.3pt} p_1 \hspace*{-0.8pt} - \hspace*{-0.8pt} \epsilon_f^{b*} \hspace*{-2pt} \cdot \hspace*{-1.3pt} p_1 \epsilon_i^b \hspace*{-2pt} \cdot \hspace*{-1.3pt} q) \nonumber \\
&& \hspace*{146pt} + \frac{1}{2} \, \frac{1}{m_b^2} \epsilon_f^{b*} \hspace*{-1pt} \cdot \hspace*{-0.5pt} q \epsilon_i^b \hspace*{-1pt} \cdot \hspace*{-0.5pt} q \Bigg) \nonumber \\
{}^1{\cal M}^{(2)}_{\ref{fig_diags}e}(q) \hspace*{-5pt} &=\hspace*{-5pt} & G \alpha Z_a Z_b \Bigg[
\frac{4 m_a m_b} {q^2} L \bigg( \hspace*{-3.5pt} - \hspace*{-1pt} 2 \epsilon_f^{*b} \hspace*{-2pt} \cdot \hspace*{-1.5pt} \epsilon_i^b
  - \hspace*{-0.7pt} \frac{3}{m_a m_b} (\epsilon_f^{b*} \hspace*{-2pt} \cdot \hspace*{-1.5pt} q \epsilon_i^b \hspace*{-2pt} \cdot \hspace*{-1.5pt} p_1 \hspace*{-1.5pt} - \hspace*{-1pt} \epsilon_f^{b*} \hspace*{-2pt} \cdot \hspace*{-1.5pt} p_1 \epsilon_i^b \hspace*{-2pt} \cdot \hspace*{-1.5pt} q) \nonumber \\
&& \hspace*{100pt} + \frac{1}{m_b^2} \epsilon_f^{b*}\cdot q \epsilon_i^b \cdot q\bigg) \nonumber \\
&&\hspace*{41pt} + \hspace*{-1pt} \bigg( \hspace*{-5pt} - \hspace*{-2pt} \frac{12 m_a^2 \hspace*{-1.5pt} + \hspace*{-1.5pt} 59 m_a m_b \hspace*{-1.5pt} + \hspace*{-1.5pt} 21 m_b^2}{3 m_a m_b} \hspace*{1pt} L \hspace*{-0.5pt} - \hspace*{-0.5pt}(3 m_a \hspace*{-1.5pt} + \hspace*{-1.5pt} 5 m_b) S \hspace*{-1pt} \bigg) \epsilon_f^{*b} \hspace*{-1.5pt} \cdot \hspace*{-1pt} \epsilon_i^b \nonumber\\
&& \hspace*{41pt} + \frac{20}{3} \hspace*{1pt} L \hspace*{1pt} \,  \frac{1}{m_a m_b} \hspace*{1pt} \epsilon_f^{b*}\cdot p_1 \epsilon_i^b \cdot p_1 \nonumber \\
&& \hspace*{41pt} + \bigg( \hspace*{-4pt} - \hspace*{-2pt} \frac{78 m_a^2 + 164 m_a m_b + 99 m_b^2}{12 m_a m_b} \hspace*{1pt} L - \frac{9}{2} \left(m_a + m_b\right) \hspace*{-1pt} S \bigg) \nonumber \\
&& \hspace*{59pt} \times \frac{1}{m_a m_b} (\epsilon_f^{b*}\cdot q \epsilon_i^b \cdot p_1 - \epsilon_f^{b*} \cdot p_1 \epsilon_i^b\cdot q) \nonumber \\
&& \hspace*{41pt} - \frac{10}{3} \hspace*{1pt} L \, \frac{1}{m_a m_b} (\epsilon_f^{b*} \hspace*{-1.5pt} \cdot \hspace*{-0.8pt} q \epsilon_i^b \hspace*{-1.5pt} \cdot \hspace*{-0.8pt} p_1 \hspace*{-1pt} + \hspace*{-0.7pt} \epsilon_f^{b*} \hspace*{-1.5pt} \cdot \hspace*{-0.8pt} p_1 \epsilon_i^b \hspace*{-1.5pt} \cdot \hspace*{-0.8pt} q)\nonumber \\
&& \hspace*{41pt} + \bigg(\hspace*{-2.5pt}  \bigg( \hspace*{-1pt} \frac{7 m_a}{3 m_b} \hspace*{-1.5pt} + \hspace*{-1.5pt} \frac{7}{6} \hspace*{-1.5pt} + \hspace*{-1.5pt} \frac{16 m_b}{3 m_a} \hspace*{-1.5pt} - \hspace*{-1.5pt} \frac{m_b^2}{4m_a^2}\bigg) \hspace*{-1pt} L \nonumber \\
&& \hspace*{55pt} + \hspace*{-1pt} \bigg(\hspace*{-1pt} \frac{5}{4} m_a \hspace*{-1.5pt} + \hspace*{-1.5pt} \frac{9}{4} m_b \hspace*{-1.5pt} - \hspace*{-1.5pt} \frac{m_b^2}{8 m_a} \hspace*{-1.5pt} \bigg)\hspace*{-1pt} S \hspace*{-1pt}  \bigg) \times \frac{1}{m_b^2} \epsilon_f^{b*}\cdot q \epsilon_i^b \cdot q  \Bigg] \nonumber \\
{}^1{\cal M}^{(2)}_{\ref{fig_diags}f_\gamma}(q) \hspace*{-5pt} &=\hspace*{-5pt} & G \alpha Z_a Z_b \Bigg[-2 \left(m_a + m_b\right) S \, \epsilon_f^{*b} \cdot \epsilon_i^b \nonumber\\
&& \hspace*{41pt} - \left(2 m_a + m_b \right) S \, \frac{1}{m_a m_b} (\epsilon_f^{b*} \hspace*{-2pt} \cdot \hspace*{-1pt} q \epsilon_i^b \hspace*{-1pt} \cdot \hspace*{-1pt} p_1 \hspace*{-1.5pt} - \hspace*{-1pt} \epsilon_f^{b*} \hspace*{-2pt} \cdot \hspace*{-1pt} p_1 \epsilon_i^b \hspace*{-1pt}\cdot \hspace*{-1pt} q) \nonumber\\
&& \hspace*{41pt} + \bigg(\hspace*{-3pt} - \frac{8}{3} L - \frac{3}{2} m_b S \bigg) \frac{1}{m_b^2} \epsilon_f^{b*}\cdot q \epsilon_i^b \cdot q \Bigg] \nonumber\\
{}^1{\cal M}^{(2)}_{\ref{fig_diags}f_g}(q) \hspace*{-5pt} &=\hspace*{-5pt} & G \alpha \Bigg[\left(\frac{8(Z_a^2 m_b^2 + Z_b^2 m_a^2)}{3 m_a m_b} \hspace*{1pt} L + \left(Z_a^2 m_b + Z_b^2 m_a\right) S \right) \epsilon_f^{*b} \cdot \epsilon_i^b \nonumber\\
&& \hspace*{41pt} + \left(\frac{4 (3Z_a^2 m_b^2 + Z_b^2 m_a^2)}{3 m_a m_b}\hspace*{1pt} L + \hspace*{-1.5pt} \left( \hspace*{-1pt} \frac{3}{2} Z_a^2 m_b \hspace*{-1pt} + \hspace*{-1pt} Z_b^2 m_a \hspace*{-1.5pt} \right)\hspace*{-2pt} S \hspace*{-1pt} \right) \nonumber \\
 &&\hspace*{59pt} \times \frac{1}{m_a m_b} (\epsilon_f^{b*} \hspace*{-2pt} \cdot \hspace*{-1pt} q \epsilon_i^b \hspace*{-1pt} \cdot \hspace*{-1pt} p_1 \hspace*{-1.5pt} - \hspace*{-1pt} \epsilon_f^{b*} \hspace*{-2pt} \cdot \hspace*{-1pt} p_1 \epsilon_i^b \hspace*{-1pt}\cdot \hspace*{-1pt} q) \nonumber\\
&& \hspace*{41pt} + \hspace*{1pt} \bigg(\frac{2(-3 Z_a^2 m_b^2 + Z_b^2 m_a^2)}{3 m_a m_b} \hspace*{1pt} L + \hspace*{-1.5pt}\left(\hspace*{-3pt} - \frac{3}{4} Z_a^2 m_b \hspace*{-1.5pt} - \hspace*{-1.5pt} \frac{1}{4} Z_b^2 m_a \hspace*{-2pt} \right) \hspace*{-1.5pt} S \hspace*{-1pt} \bigg)  \nonumber \\
&& \hspace*{59pt} \times \frac{1}{m_b^2} \epsilon_f^{b*}\cdot q \epsilon_i^b \cdot q \Bigg] \nonumber\\
{}^1{\cal M}^{(2)}_{\ref{fig_diags}g}(q) \hspace*{-5pt} &=\hspace*{-5pt} & G \alpha Z_a Z_b \Bigg[\hspace*{-1.5pt} - \frac{4}{3} \hspace*{1pt} L \hspace*{2pt} \epsilon_f^{*b} \cdot \epsilon_i^b \nonumber \\
&& \hspace*{41pt} - \frac{4}{3} \hspace*{1pt} L \hspace*{1pt} \frac{1}{m_a m_b} (\epsilon_f^{b*}\cdot q \epsilon_i^b \cdot p_1 - \epsilon_f^{b*} \cdot p_1 \epsilon_i^b\cdot q) \Bigg]. \label{eq:lh}
\end{eqnarray}
where $\epsilon^b_{i\mu}$ is the polarization
vector for the incoming spin 1 particle and $\epsilon^b_{f\mu}$
is its outgoing polarization vector which satisfy
$\epsilon_f^{*b}\cdot p_4=0$ and $\epsilon_i^{b}\cdot p_3=0$
respectively.

It is useful at this point to define the spin four-vector
\begin{equation}
S_{b\mu}={i\over
2m_b}\epsilon_{\mu\beta\gamma\delta}\epsilon_f^{b*\beta}\epsilon_i^{b\gamma}(p_3+p_4)^\delta
\end{equation}
which satisfies the identity
\begin{equation}
\epsilon_{f\mu}^{*b}\epsilon_i^b\cdot
q-\epsilon_{i\mu}^b\epsilon_f^{*b}\cdot q={1\over 1-{q^2\over
4m_b^2}}\left[{i\over m_b}\epsilon_{\mu\beta\gamma\delta}p_3^\beta
q^\gamma S_b^\delta-{(p_3+p_4)_\mu\over 2m_b^2}\epsilon_f^{*b}\cdot
q\epsilon_i^b\cdot q\right]\label{eq:df}
\end{equation}
Using this definition of the spin vector and the identity Eq.
(\ref{eq:df}) we find that the sum of the diagrams Eq. (\ref{eq:lh}) has
the form
\begin{eqnarray}
{}^1{\cal M}_{tot}^{(2)}(q)\hspace*{-5pt}&=&\hspace*{-5pt} G \alpha Z_a Z_b \Bigg[ \hspace*{-1.5pt} - \epsilon_f^{b*} \cdot \epsilon_i^b \hspace*{1pt} \Big( \hspace*{-1pt} -6 (m_a+m_b)S + 12 L \Big) \nonumber \\
&& \hspace*{34pt} +  i \frac{{\mathcal E}_b}{m_a m_b^2} \hspace*{-2.7pt} \left( \hspace*{-4.5pt}\left( \hspace*{-2.8pt} -\frac{2 m_a m_b (\hspace*{-0.4pt} 2 m_a \hspace*{-2.7pt} + \hspace*{-2.2pt} 3 m_b \hspace*{-0.4pt} )}{s - s_0} \hspace*{-1.5pt} - \hspace*{-1.9pt} 5 m_a \hspace*{-2.7pt} - \hspace*{-1.7pt} \frac{15}{2} m_b \hspace*{-3.4pt}\right) \hspace*{-2.5pt} S \hspace*{-1.8pt} + \hspace*{-1.7pt} \frac{32}{3} L \hspace*{-2.3pt}\right) \hspace*{-2pt}\nonumber \\
&& \hspace*{34pt} + \frac{\epsilon_f^{b*} \hspace*{-1.5pt} \cdot \hspace*{-1.5pt} q\epsilon_i^b \hspace*{-1.5pt} \cdot \hspace*{-1.5pt} q}{m_b^2} \hspace*{-2.8pt} \left( \hspace*{-4.5pt}\left( \hspace*{-1pt} \frac{m_a m_b (m_a \hspace*{-2.5pt} + \hspace*{-2.2pt} m_b)}{s - s_0} \hspace*{-1.3pt} + \hspace*{-1.3pt} \frac{25}{4} m_a \hspace*{-2.7pt} + \hspace*{-1.3pt} \frac{35}{4} m_b \hspace*{-3.4pt}\right) \hspace*{-2.5pt} S \hspace*{-1.8pt} - \hspace*{-2pt} \frac{28}{3} L \hspace*{-3pt}\right) \hspace*{-4.4pt} \Bigg] \nonumber \\
& + & \hspace*{-5pt} G \alpha \Bigg[\hspace*{-1.5pt} - \epsilon_f^{b*} \cdot \epsilon_i^b \left( - (Z_a^2 m_b + Z_b^2 m_a) S - \frac{8(Z_a^2 m_b^2 + Z_b^2 m_a^2)}{3 m_a m_b} \hspace*{1pt} L \right) \nonumber \\
 && \hspace*{10pt} + i \frac{{\mathcal E}_b}{m_a m_b^2} \bigg(\hspace*{-3pt} - \hspace*{-1pt} \left(\frac{3}{2} Z_a^2 m_b + Z_b^2 m_a \hspace*{-1pt}\right) \hspace*{-1pt} S - \frac{4 (3 Z_a^2 m_b^2 + Z_b^2 m_a^2)}{3 m_a m_b} \hspace*{1pt} L \hspace*{-1pt}\bigg) \hspace*{-1pt} \nonumber \\
&& \hspace*{10pt} + \frac{\epsilon_f^{b*} \hspace*{-1pt} \cdot \hspace*{-1pt} q\epsilon_i^b \hspace*{-1pt} \cdot \hspace*{-1pt} q}{m_b^2} \hspace*{-1pt} \left(\frac{3}{4}\left(Z_a^2 m_b + Z_b^2 m_a\right)S + \frac{2(Z_a^2 m_b^2 + Z_b^2 m_a^2)}{m_a m_b} \hspace*{1pt} L\right) \hspace*{-3pt} \Bigg] \nonumber \\
&+& \hspace*{-5pt} i 8\pi G  m_a m_b \hspace*{1.3pt} \alpha Z_a Z_b \hspace*{1.5pt} {L\over q^2} \sqrt{m_a m_b \over s-s_0} \hspace*{-1pt} \left(\hspace*{-2pt} - \epsilon_f^{b*} \hspace*{-2pt} \cdot \hspace*{-2pt} \epsilon_i^b \hspace*{-1.5pt} + \hspace*{-1.5pt} \frac{3}{2} \hspace*{1.5pt} i \frac{{\mathcal E}_b}{m_a m_b^2} \hspace*{-0.9pt} - \hspace*{-0.9pt} \frac{\epsilon_f^{b*} \hspace*{-2pt} \cdot \hspace*{-2pt} q\epsilon_i^b \hspace*{-2pt} \cdot \hspace*{-2pt} q}{m_b^2}\hspace*{-1pt} \right) \hspace*{-2pt}. \nonumber \\ \label{eq:os}
\quad
\end{eqnarray}

In order to generate a potential, we must first perform the
nonrelativistic limit, wherein
\begin{equation}
\epsilon_i^{b0}\simeq {1\over m_b} \hspace*{1pt}
\hat{\epsilon}_i^b\cdot\vec{p_3},\quad\epsilon_f^{b0}\simeq {1\over
m_b} \hspace*{1pt} \hat{\epsilon}_f^b\cdot\vec{p_4}
\end{equation}
so that
\begin{eqnarray}
\epsilon_f^{b*}\cdot\epsilon_i^b&\simeq&
-\hat{\epsilon}_f^{b*}\cdot\hat{\epsilon}_i^b+{1\over
m_b^2} \hspace*{1pt} \hat{\epsilon}_f^{b*}\cdot\vec{p}_4 \hspace*{1pt} \hat{\epsilon}_i^b\cdot\vec{p}_3\nonumber\\
&\simeq&-\hat{\epsilon}_f^{b*}\cdot\hat{\epsilon}_i^b+{1\over
2m_b^2}\hat{\epsilon}_f^{b*}\times\hat{\epsilon}_i^b\cdot\vec{p}_4\times\vec{p}_3\nonumber\\
&&+{1\over
2m_b^2}\left(\hat{\epsilon}_f^{b*}\cdot\vec{p}_4\hat{\epsilon}_i^b\cdot\vec{p}_3
+\hat{\epsilon}_f^{b*}\cdot\vec{p}_3\hat{\epsilon}_i^b\cdot\vec{p}_4\right).
\label{eq:sn}
\end{eqnarray}
Since
\begin{equation}
-i\hat{\epsilon}_f^{b*}\times\hat{\epsilon}_i^b= \left<1,m_f \left|
\vec{S}_b \right| 1,m_i\right>,
\end{equation}
Eq. (\ref{eq:sn}) becomes
\begin{eqnarray}
\epsilon_f^{b*}\cdot\epsilon_i^b \hspace*{-0.2pt}{}& \simeq&
-\hat{\epsilon}_f^{b*}\cdot\hat{\epsilon}_i^b-{i\over 2m_b^2}
\hspace*{1pt} \vec{S}_b\cdot\vec{p}_3\times\vec{p}_4 +{1\over
2m_b^2} \! \hspace*{-0.2pt}
\left(\hat{\epsilon}_f^{b*}\cdot\vec{p}_4 \hspace*{1pt}
\hat{\epsilon}_i^b\cdot\vec{p}_3
+\hat{\epsilon}_f^{b*}\cdot\vec{p}_3 \hspace*{1pt} \hat{\epsilon}_i^b\cdot\vec{p}_4\right) \nonumber \\
& \simeq& -\hat{\epsilon}_f^{b*}\cdot\hat{\epsilon}_i^b +{1\over
m_b^2} \hspace*{1pt} \hat{\epsilon}_f^{b*}\cdot\vec{p} \
\hat{\epsilon}_i^b\cdot\vec{p} +{i \over 2m_b^2} \hspace*{1pt}
\vec{S}_b\cdot\vec{p} \times\vec{q} -{1\over 4 m_b^2} \hspace*{1pt}
\hat{\epsilon}_f^{b*}\cdot\vec{q} \ \hat{\epsilon}_i^b\cdot\vec{q}.
\nonumber \\
\end{eqnarray}
Comparing with the corresponding nonrelativistic reduction for the spin-1/2 particle
in Eq. (\ref{eq:nr}) we see that the structure of the spin-independent and spin-orbit
pieces of $-\epsilon_f^{b*}\cdot\epsilon_i^b $ is identical to that of $\bar{u}(p_4)u(p_3)$.
However, in the unit spin case these terms are accompanied by new terms which are
quadrupole in nature, as can see from the identity
\begin{eqnarray}
T^b_{cd}&\equiv&{1\over 2}\left(\hat{\epsilon}_{fc}^{b*}
\hspace*{1pt}\hat{\epsilon}_{id}^b+ \hat{\epsilon}_{ic}^b
\hspace*{1pt}\hat{\epsilon}_{fd}^{b*}\right)-{1\over
3} \hspace*{1pt} \delta_{cd} \hspace*{1pt} \hat{\epsilon}_f^{b*}\cdot\hat{\epsilon}_i^b\nonumber\\
&=& - \left<1,m_f \left|{1\over 2} (S_cS_d+S_dS_c)-{2\over
3}\delta_{cd} \right|1,m_i \right>
\end{eqnarray}
Finally, comparing the total amplitude for spin-0 -- spin-1 scattering
Eq. (\ref{eq:os}) and for spin-0 -- spin-1/2 scattering Eq.
(\ref{eq_mix_Mh_rel}) we observe that the spin-independent and
spin-orbit pieces are identical once we replace the spin-1/2
structure $\bar{u}(p_4)u(p_3)$ by
$-\epsilon_f^{*b}\cdot\epsilon_i^b$ in the spin-1 case.  Since the
nonrelativistic reductions of these structures are also the same up
to quadrupole corrections---
\begin{eqnarray}
\bar{u}(p_4)u(p_3) & \stackrel{NR}{\longrightarrow} &
\chi_f^{b\dagger}\chi_i^b - {i\over 2m_b^2}\vec{S}_b\cdot \vec{p} \times\vec{q} \nonumber \\
-\epsilon_f^{*b}\cdot\epsilon_i^b & \stackrel{NR}{\longrightarrow} &
\hat{\epsilon}_f^{b*}\cdot\hat{\epsilon}_i^b - {i\over 2m_b^2}\vec{S}_b\cdot \vec{p} \times\vec{q} +\ldots
\end{eqnarray}
where $\chi_f^{b\dagger}\chi_i^b = \delta_{m_f^b, m_i^b}$ and
$\hat{\epsilon}_f^{b*}\cdot\hat{\epsilon}_i^b = \delta_{m_f^b,
m_i^b}$--- we observe that the spin-independent and spin-orbit
pieces of the nonrelativistic reductions of the spin-0 -- spin-1/2 and
spin-0 -- spin-1 amplitudes are also identical.  Finally, since the
lowest order electromagnetic and gravitational amplitudes/potentials
for spin-0 -- spin-1 scattering have the same forms for the $S-I$ and
$S-O$ components as their spin-0 -- spin-1/2 counterparts \cite{hrem,
hrgr}, it is clear that the second order potential for spin-0 --
spin-1 scattering is identical to its spin-0 -- spin-0 (for $S-I$) and
both spin-0 -- spin-1/2 and spin-1/2 -- spin-1/2 (for both $S-I$ and
$S-O$) counterparts. However, clearly there are new terms in the
potential which are quadrupole in nature and have no counterpart in
lower spin systems. The form of the quadrupole potentials is more
complex and is currently under study.

\section{Conclusions}

Above we have presented a series of calculations of mixed
electromagnetic-gravitational ({\it i.e.}, $\mathcal O(G \alpha)$)
effects in the long distance piece of the scattering
amplitude/potential of two massive particles for various spin
combinations.  The basic idea is to use the methods of effective
field theory for such a process, as outlined by Donoghue for the
related process of second order gravitational scattering \cite{don}.
The pieces of the scattering amplitude which lead to power law
behavior of the coordinate space potential are nonanalytic in
character and are of two forms. One involves terms behaving as $1 /
\sqrt{-q^2}$ and the second involves nonanalytic pieces having the
form $\log -q^2$. When Fourier transformed the first form leads to
terms which are classical ($\hbar$-independent) and which lead to
long distance spin-independent behavior $V_{class}(r)\sim G\alpha m
/ r^2$, while the second one is quantum mechanical
($\hbar$-dependent) and involves pieces of the potential which have
the form $V_{qm}(r)\sim \hbar G \alpha/ r^3$. The project was
undertaken both because of the intrinsic interest of such a
calculation but also in order to resolve the disagreement between
various papers on this subject which have recently appeared
\cite{bjb, msb, svf}.  Specifically, the existing publications have
{\it not} satisfied the universality property, which asserts that
the kinematic forms for various pieces of the scattering amplitude
should be identical, regardless of the spin-content of the
scattering particles.  By carefully evaluating the scattering of
particles with spins 0-0, 0-1/2, 0-1, and 1/2-1/2 we demonstrated
that the results of all existing publications were incorrect (note,
however, that the simple typo in the work of Bjerrum-Bohr has now
been corrected \cite{nb2}), and that universality {\it is}
satisfied.

Our results are presented in the form of a second order
potential, which is defined in terms of the Fourier transform
\begin{equation}
V^{(2)}_{CG}(\vec{r})=-\int {d^3q\over
(2\pi)^3}e^{-i\vec{q}\cdot\vec{r}} \left[{\cal
M}^{(2)}_{tot}(\vec{q})-{\rm Amp}_{CG}^{(2)}(\vec{q})\right]\label{eq:bo}
\end{equation}
where ${\cal M}^{(2)}_{tot}(\vec{q})$ is the full second order
scattering amplitude, which includes all one loop amplitudes
involving both graviton and photon exchange, and ${\rm
Amp}^{(2)}(\vec{q})$ is the second Born approximation amplitude to
the same process.  The subtraction of the iterated Born piece is
necessary in order to eliminate pieces of the second order
scattering amplitude which diverge as the center of mass momentum
$p_0$ approaches zero.  These terms are of two sorts:
\begin{itemize}
\item [i)] imaginary components which behave as $i \hspace*{0.6pt} \log - q^2 / (q^2 p_0)$
\item [ii)] real pieces which behave as $1 / (\sqrt{-q^2} p_0^2)$
\end{itemize}
Both such terms were shown to disappear when the subtraction shown
in Eq. (\ref{eq:bo}) is performed.  However, it should be noted that
while the quantum mechanical component of the second order
potential is unique, the classical component of the potential
contains an ambiguity in that, as pointed out by Sucher \cite{jsu}, the
iterated Born amplitude depends the choice for the lowest order
potential (see also \cite{hrem}). Moreover, due to general covariance
the classical component also depends on the choice of coordinates \cite{bjb, BjerrumBohr:2002kt}.
The subtraction of the offending
forms i) and ii) above is independent of these choices, but what
remains behind are finite pieces proportional to $1/\sqrt{-q^2}$
whose coefficient depends upon this choice so that the
resulting $\mathcal O(G \alpha)$ classical potential is not unique.
Of course, this ambiguity is not of any physical significance, since the
potential itself is not an observable. Rather the only observable is the
second order on-shell scattering amplitude which is invariant and well
defined.

Our results for the second order potential can be summarized
succinctly---for arbitrary spin scattering there exists a
spin-independent component which has the form
\begin{eqnarray}
 V_{S-I}^{(2)}(\vec{r}) \hspace*{-6pt}& \simeq \hspace*{-6pt} & G \alpha \Bigg[\frac{1}{2} \frac{ (Z_a^2 m_b + Z_b^2 m_a)}{r^2} + 3 \frac{Z_a Z_b (m_a + m_b)}{r^2} \nonumber \\
  && {} \hspace*{15pt} - \frac{4 \hbar}{{3 \pi r^3}} \left(Z_b^2 \frac{m_a}{m_b} + Z_a^2 \frac{m_b}{m_a}\right) + \frac{6 Z_a Z_b \hbar}{\pi r^3} \Bigg] \, \delta_{m_f^b, m_i^b}
\end{eqnarray}
where $m^{a,b}_{i,f}$ represents the projection on the quantization
axis of the spin of the indicated particle.  In addition, if either
particle carries spin there exists an additional shorter-range
spin-orbit contribution to the potential which is seen in Eqs.
(\ref{eq:hd}) and (\ref{eq_nlo_mixhh}) and whose universality is
shown via 0-1/2, 0-1 and 1/2-1/2 scattering. One subtlety here is
that the spin-orbit coupling does depend on the g-factor of the
scattered particles with spin, and we have used the "natural" value
$g=2$ throughout. See \cite{hrem} for a discussion on the dependence
of the spin-dependent terms on $g$ in purely electromagnetic
scattering. If both particles carry spin a new spin-spin coupling
arises as seen in Eq. (\ref{eq_nlo_mixhh}) which is even shorter
ranged than the spin-orbit coupling and whose universality can also
be presumed. Finally, if one of the spins is one or greater there
exist also quadrupole components of the potential which are even
shorter range and (presumably) universal.  These forms are still
under study.

In any case, we have demonstrated that universality is valid for the
long distance components in the case of mixed electromagnetic-gravitational
scattering and that previous indications to the contrary were due to calculational
errors.
\begin{center}
{\bf Acknowledgements}
\end{center}

We thank John Donoghue for discussions about this calculation.  This
work was supported in part by the National Science Foundation under
award PHY05-53304 (BRH and AR) and by the the US Department of Energy under
grant DE-FG-02-92ER40704 (AR).

\appendix

\section{Feynman Rules} \label{app_fmr}

In order to carry out the calculations described in the text we
require the appropriate electromagnetic, gravitational, and mixed
vertices. The purely electromagnetic vertices which involve photons
but no gravitons are found in \cite{hrem}, and the purely gravitational
vertices involving gravitons and no photons are found in \cite{hrgr}.
Here we only list the vertices with couplings to both photons and
gravitons which are specific to this work and which were not needed
in \cite{hrem} and \cite{hrgr}.

For all couplings involving photons we use a g-factor of $g=2$ at tree
level. The g-factor does not affect the spin-independent components
of our results, but the spin-dependent ones do depend on it. In \cite{hrem}
we offer a more detailed discussion and calculate purely electromagnetic
scattering for arbitrary g-factors in an appendix.

\subsection{Photon-Graviton Coupling}

When minimally coupling the Maxwell Lagrangian
\begin{equation}
{\cal L}= - \frac{1}{4} \hspace*{1pt} F_{\mu \nu} F^{\mu \nu} \label{eq_maxwell_lagr}
\end{equation}
with $F_{\mu \nu} = \partial_\mu A_\nu - \partial_\nu A_\mu$
to gravity, a vertex arises which couples a graviton to two photons. Additional contributions to this
vertex arise from the electromagnetic gauge fixing Lagrangian
\begin{equation}
 {\cal L}_{GF} = - \frac{1}{2} (\partial_\mu A^\mu)^2 \label{eq_em_gf_lagr}
\end{equation}
coupled to gravity which have to be included in order to be consistent. Our vertex reads
\begin{figure}[!h]
  \centering
  \includegraphics{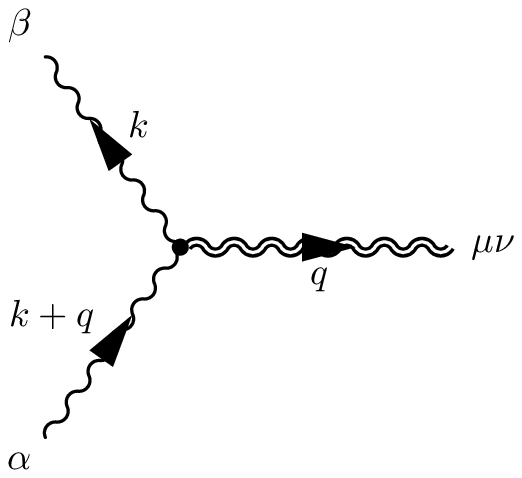}
\end{figure}
\begin{eqnarray}
 \tau_{\alpha,\beta,\mu \nu}(k,q) \hspace*{-6pt} & = & \hspace*{-6pt} \frac{i \kappa}{2} \Big[2 \hspace*{0.7pt} P_{\mu \nu, \alpha \beta} \, k \hspace*{-1.5pt} \cdot \hspace*{-1.5pt} (k \hspace*{-1pt} + \hspace*{-1pt} q) \hspace*{-1pt} + \hspace*{-1pt} \eta_{\mu \nu} k_\alpha (k \hspace*{-1pt} + \hspace*{-1pt} q)_\beta \hspace*{-1pt} \nonumber \\
 && \hspace*{6pt} + \hspace*{-0.5pt} \eta_{\alpha \beta} \big(k_\mu (k \hspace*{-1pt} + \hspace*{-1pt} q)_\nu \hspace*{-1pt} + \hspace*{-1pt} (k \hspace*{-1pt} + \hspace*{-1pt} q)_\mu k_\nu \big) \nonumber \\
 && \hspace*{6pt} - \hspace*{0.5pt} \eta_{\mu \alpha} k_\nu (k \hspace*{-1pt} + \hspace*{-1pt} q)_\beta \hspace*{-1pt}
                  - \hspace*{-1pt} \eta_{\mu \beta} k_\alpha (k \hspace*{-1pt} + \hspace*{-1pt} q)_\nu \hspace*{-1pt} \nonumber \\
 && \hspace*{6pt} - \hspace*{-0.5pt} \eta_{\nu \alpha} k_\mu (k \hspace*{-1pt} + \hspace*{-1pt} q)_\beta \hspace*{-1pt}
                 - \hspace*{-1pt} \eta_{\nu \beta} k_\alpha (k \hspace*{-1pt} + \hspace*{-1pt} q)_\mu \Big]\nonumber \\
 & + & \hspace*{-6pt} \frac{i \kappa}{2} \bigg[\eta_{\mu \alpha} k_\nu k_\beta \hspace*{-1pt}
                 + \hspace*{-1pt} \eta_{\mu \beta} (k \hspace*{-1pt} + \hspace*{-1pt} q)_\alpha (k \hspace*{-1pt} + \hspace*{-1pt} q)_\nu \hspace*{-1pt} \nonumber \\
 && \hspace*{6pt} + \hspace*{-0.5pt} \eta_{\nu \alpha} k_\mu k_\beta \hspace*{-1pt}
                 + \hspace*{-1pt} \eta_{\nu \beta} (k \hspace*{-1pt} + \hspace*{-1pt} q)_\alpha (k \hspace*{-1pt} + \hspace*{-1pt} q)_\mu \nonumber \\
 && \hspace*{6pt} +\eta_{\mu \nu} \Big((k \hspace*{-1pt} + \hspace*{-1pt} q)_\alpha k_\beta - k_\alpha k_\beta - (k \hspace*{-1pt} + \hspace*{-1pt} q)_\alpha (k \hspace*{-1pt} + \hspace*{-1pt} q)_\beta \hspace*{-1pt}\Big) \bigg]
\end{eqnarray}
with
$P_{\alpha\beta,\gamma\delta} \equiv {1\over 2}(\eta_{\alpha\gamma}\eta_{\beta\delta}+\eta_{\alpha\delta}\eta_{\beta\gamma} - \eta_{\alpha\beta}\eta_{\gamma\delta})$
where the first term stems from the Maxwell Lagrangian coupled to gravity and the second term is derived from the electromagnetic
gauge fixing Lagrangian coupled to gravity. The gauge fixing contributions to the vertex were omitted in previous publications
\cite{Donoghue:2001qc, bjb, BjerrumBohr:2004mz, svf}. Even though they change individual
diagrams that contribute to form factors, they do not seem to affect any physical quantities so that the total results of
the work in \cite{Donoghue:2001qc, bjb, BjerrumBohr:2004mz} are correct (up to the typo in \cite{bjb}
pointed out above which is also found in \cite{BjerrumBohr:2004mz}).

\newpage

\subsection{Couplings to Spin-0 Particles}
The Feynman rules for a scalar particle with both electromagnetic and gravitational interactions can be derived by coupling both
interactions minimally to the Klein-Gordon Lagrangian. The relevant vertex with one photon and one graviton coupled
is found to be
\begin{figure}[h]
  \centering
  \includegraphics{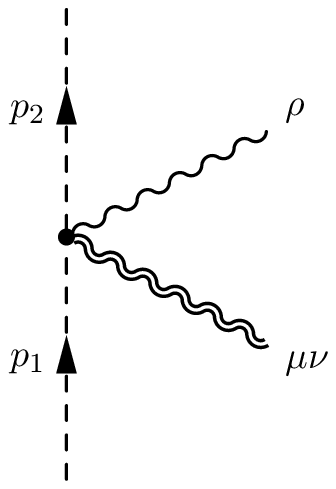}
\end{figure}
\vspace*{-25pt}
\begin{eqnarray}
 {} \hspace*{21pt} {}^0 \tau_{\mu \nu, \rho}^{(1,1)} \hspace*{-5pt} & = & \hspace*{-5pt} i Z e \hspace*{1pt} \kappa P_{\mu \nu, \rho \lambda} \hspace*{1pt} (p_1+p_2)^\lambda
\end{eqnarray}
where the charge of the spin-0 particle is $Z e$.

\subsection{Couplings to Spin-1/2 Particles}
Similarly, the mixed vertex for a spin-1/2 particle with g-factor $g=2$ coupled to both a graviton and a photon is
\begin{figure}[h]
  \centering
  \includegraphics{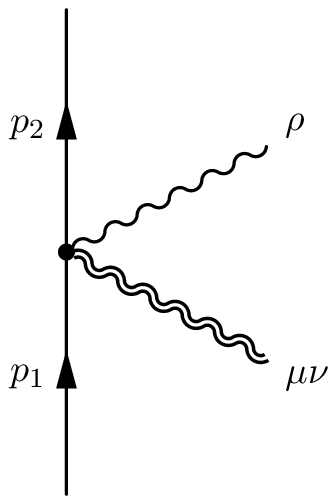}
\end{figure}
\vspace*{-25pt}
\begin{eqnarray}
 {} \hspace*{83pt} {}^{\frac{1}{2}} \tau_{\mu \nu, \rho}^{(1,1)} \hspace*{-5pt} & = & \hspace*{-5pt} - \frac{i Z e \hspace*{1pt} \kappa}{4} \Big(2 \eta_{\mu \nu} \gamma_\rho - \eta_{\mu \rho} \gamma_\nu - \eta_{\nu \rho} \gamma_\mu \Big).
\end{eqnarray}

\subsection{Couplings to Spin-1 Particles}
The mixed vertex for a spin-1 particle with g-factor $g=2$ at tree level coupled to both a graviton and a photon
has the expression
\begin{figure}[h]
  \centering
  \includegraphics{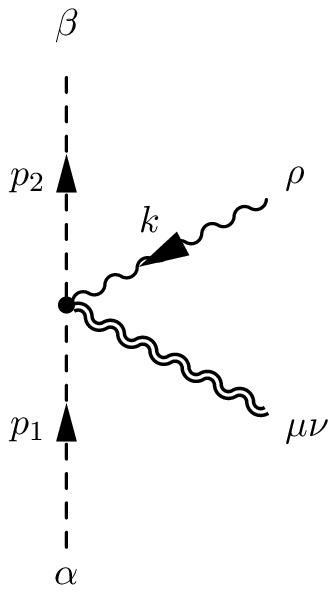}
\end{figure}
\vspace*{-20pt}
\begin{eqnarray}
 {} \hspace*{114pt} {}^1 \tau_{\beta, \alpha, \mu \nu, \rho}^{(1,1)} \hspace*{-5pt} & = & \hspace*{-5pt} - i Z e \hspace*{1pt} \kappa \Big((p_1 + p_2)^\lambda \big(I_{\alpha \beta, \mu \nu} \eta_{\rho \lambda} + \eta_{\alpha \beta} P_{\mu \nu, \rho \lambda}\big) \nonumber \\
  && \hspace*{29pt} - (p_1 \hspace*{-0.2pt} - \hspace*{-0.2pt} k)^\lambda \big(I_{\alpha \rho, \mu \nu} \eta_{\beta \lambda} \hspace*{-0.2pt} + \hspace*{-0.2pt} \eta_{\alpha \rho} P_{\mu \nu, \beta \lambda}\big) \nonumber \\
  && \hspace*{29pt} - (p_2 \hspace*{-0.2pt} + \hspace*{-0.2pt} k)^\lambda \big(I_{\alpha \lambda, \mu \nu} \eta_{\beta \rho} \hspace*{-0.2pt} + \hspace*{-0.2pt} \eta_{\alpha \lambda} P_{\mu \nu, \beta \rho}\big) \Big). \nonumber \\ \quad
\end{eqnarray}

\section{Iteration Integrals}

In this appendix we give the integrals
\begin{equation}
[H;H_r;H_{rs}]=i \int {d^3\ell\over (2\pi)^3}
{- 4\pi Gm_am_b \over
|\vec{\ell} - \vec{p}_f|^2+\lambda^2}{i[1;\ell_r;\ell_r\ell_s]\over
{p_0^2\over 2m_r}-{\ell^2\over 2m_r}+i\epsilon}{4 \pi \alpha Z_aZ_b \over
|\vec{p}_i - \vec{\ell}|^2+\lambda^2}
\end{equation}
which are needed in order to perform the iteration of the lowest
order Newton potentials. Here we list only the results; for a
more detailed derivation, albeit with a different prefactor, see
\cite{hrem}. The leading expressions for the iteration integrals read
\begin{eqnarray}
H &\simeq&i \hspace*{0.3pt} 4\pi G m_am_b \, \alpha Z_aZ_b \hspace*{1pt} \frac{L}{q^2} \frac{m_r}{p_0}\nonumber\\
H_r &\simeq&(p_i+p_f)_r \ G m_am_b \, \alpha Z_aZ_b
\left( i \hspace*{0.5pt} 2\pi \frac{L}{q^2} \frac{m_r}{p_0}- S {m_r\over p_0^2}+\ldots\right)\nonumber\\
H_{rs} &\simeq& \delta_{rs} \ \vec
q^{\hspace*{1.4pt}2} \ G m_am_b \, \alpha Z_aZ_b
\left(-i\pi \frac{L}{q^2} \frac{m_r}{p_0} + \frac{1}{2} S{m_r\over p_0^2}+\ldots\right) \nonumber\\
&+&(p_i+p_f)_r(p_i+p_f)_s \ G m_am_b \, \alpha Z_aZ_b \left( i\pi \frac{L}{q^2} \frac{m_r}{p_0} -  S
{m_r\over p_0^2
}+\ldots\right) \nonumber\\
&+& (p_i-p_f)_r(p_i-p_f)_s \ G m_am_b \, \alpha Z_aZ_b
\left(i\pi \frac{L}{q^2} \frac{m_r}{p_0} - \frac{1}{2} S {m_r\over p_0^2}+\ldots\right). \nonumber \\ \quad
\end{eqnarray}

\end{document}